\documentclass[journal]{IEEEtran}
\usepackage{amsthm, amssymb, amsfonts,algorithm, algpseudocode, array}
\usepackage{pifont}
\usepackage[cmex10]{amsmath}
\usepackage{cancel,cite,color,comment,dblfloatfix}
\usepackage{enumerate,esint}
\usepackage{float}
\usepackage[T1]{fontenc}
\usepackage{graphicx}
\usepackage[latin9]{inputenc}
\usepackage{multirow,multicol}
\usepackage{tikz,tgtermes}
\usepackage[labelformat=simple]{subcaption}
\usepackage{pgfplots}
\usepackage{bbm}
\usepackage{dsfont}
\usepackage{hyperref}
\usepackage{caption}
\usepackage{adjustbox}

\usetikzlibrary{backgrounds}
\usetikzlibrary{shapes, snakes}
\pgfplotsset{compat=1.11}
\allowdisplaybreaks
\newcommand{\cmark}{\ding{51}}
\newcommand{\xmark}{\ding{55}}

\theoremstyle{plain}
\newtheorem{thm}{Theorem}
\newtheorem{lem}[thm]{Lemma}

\theoremstyle{definition}
\newtheorem{defn}{Definition}

\theoremstyle{remark}

\begin{document}

\title{A family of graph GOSPA metrics for graphs with different sizes}

\author{Jinhao Gu, \'Angel F. Garc\'ia-Fern\'andez, Robert E. Firth, Lennart Svensson \thanks{J. Gu is with the Department of Electrical Engineering and Electronics, University of Liverpool, Liverpool L69 3GJ, United Kingdom (email: jinhgu@icloud.com). A. F. Garc\'ia-Fern\'andez is with the Information Processing and Telecommunications Center, ETSI de Telecomunicaci\'on, Universidad Polit\'ecnica de Madrid, 28040 Madrid, Spain (email: \mbox{angel.garcia.fernandez@upm.es}). R. E. Firth is with the STFC Hartree Centre, WA4 4AD Daresbury, United Kingdom (email: robert.firth@stfc.ac.uk). L. Svensson is with the Department of Electrical Engineering, Chalmers University of Technology, SE-412 96 Gothenburg, Sweden (email: lennart.svensson@chalmers.se). This work was supported by the EPSRC Centre for Doctoral Training in Distributed Algorithms EP/S023445/1. A. F. Garc\'ia-Fern\'andez acknowledges the research environment provided by the ELLIS Unit Madrid.
This paper has supplementary downloadable material available at \url{http://ieeexplore.ieee.org.}, provided by the authors. The material includes the appendices.}}

\maketitle
\thispagestyle{empty}
\begin{abstract}
This paper proposes a family of graph metrics for measuring distances between graphs of different sizes. The proposed metric family defines a general form of the graph generalised optimal sub-pattern assignment (GOSPA) metric and is also proved to satisfy the metric properties. Similarly to the graph GOSPA metric, the proposed graph GOSPA metric family also penalises the node attribute costs for assigned nodes between the two graphs, and the number of unassigned nodes. However, the proposed family of metrics provides more general penalties for edge mismatches than the graph GOSPA metric. This paper also shows that the graph GOSPA metric family can be approximately computed using linear programming. Simulation experiments are performed to illustrate the characteristics of the proposed graph GOSPA metric family with different choices of hyperparameters. The benefits of the proposed graph GOSPA metric family for classification tasks are also shown on real-world datasets.

\end{abstract}


\section{Introduction}

Graphs, as representations with both nodes and node connection information, are commonly used in several applications, including signal processing \cite{ekambaram15}, \cite{venkit19}, recommendation systems \cite{wang18}, chemical compound classification \cite{Smalter10}, and graph neural networks \cite{scarselli2008graph}. In applications where we infer or learn graph information, it is important to have a notion of error between graphs. We consider there is a ground truth graph, and an estimated graph, and we want to know how close the estimate is to the ground truth.

There are various methods for measuring the discrepancies among graphs and defining distance functions. For nodes without attributes, for instance, one can use the Gromov-Wasserstein discrepancy \cite[Eq. (1)]{Xu19} or the Wasserstein distance of the induced Gaussian distributions of the graph Laplacian matrices \cite{Maretic19}. The Gromov-Wasserstein discrepancy can also be extended to incorporate node attribute information, for instance, by removing the optimisation over node embeddings in \cite[Eq. (2)]{Xu19} and using the node attributes instead of the embeddings.

For a principled notion of discrepancy, it is important that the distance function is a mathematically defined metric, which meets the identity, symmetry and triangle inequality properties\footnote{In this paper, the term `distance function' refers to a function that does not necessarily meet the metric properties, while the term `metric' is reserved for a function that meets the required mathematical properties of metrics.} \cite{Apostol_book74}. In addition, it is relevant that the distance function has a clear interpretation in terms of what it penalises.

We proceed to review several metrics for graphs with interpretable results. A graph distance based on graph isomorphisms is the maximum common subgraph (MCS) distance \cite{bunke1997relation,bunke1998graph}, which computes the maximum common subgraph between two graphs. The MCS does not consider the attributes in the vertices. This implies the MCS distance does not satisfy the metric property in the space of graphs with vertex attributes.

Another commonly used method is the graph edit distance (GED) \cite{sanfeliu1983}. The GED computes the minimum edit cost of converting one graph to another graph, which has clear interpretability. The GED depends on the cost selection for each operation, which can be an insertion, deletion or substitution of a node or an edge. If these cost functions for each operation are metrics, the GED can also be a metric~\cite{Justice06}. The exact solution of the GEDs is generally NP-hard, but \cite{fischer2015approximation} and \cite{saadah2020} show that the GED can be calculated by approximate techniques to obtain polynomial computational complexity, although the approximations based on bipartite graph matching \cite{fischer2015approximation} and greedy algorithms \cite{saadah2020} do not preserve the metric property.  

In \cite{kvasnivcka1991reaction}, a distance originally designed for graphs of chemicals was introduced, but it did not consider node attributes. This chemical distance has been extended to a family of graph distances for attributed graphs in \cite{Bento19}. The family in \cite{Bento19}, which we refer to as the generalised
chemical distances (GCDs), can also include node attributes, but their mathematical formulation is limited to graphs of the same size. This approach has shown its computational benefits for large graphs in \cite{Moharrer20} with an implementation based on the alternating direction method of multipliers (ADMM) \cite{Boyd11}. In practice, GCDs can be used to compare graphs of different sizes by adding dummy nodes to the smaller graph to match the size of the larger graph. However, adding dummy nodes can break the triangle inequality property. For instance, this happens if we add dummy nodes with zeros as their attributes with no edges to other nodes, as done in \cite{Moharrer20}.

Recently, two types of graph metrics for undirected graphs (without self-loops) with a general edge attribute space have been introduced in \cite{schuhmacher2023assignment}. The graph metrics of the first type in \cite{schuhmacher2023assignment} are based on the optimal sub-pattern assignment (OSPA) metric \cite{schuhmacher2008new}. The OSPA-based metrics include a normalisation factor and associate all nodes in the smaller graph to nodes in the larger graph, penalising their node attribute errors, and including a penalty for node cardinality mismatch. However, it is possible to add extra false nodes (nodes whose attributes are sufficiently different from all the node attributes in the ground truth graph) without penalty \cite[Examples 1 and 2]{Rahmathullah17}. This property may be desirable or not, depending on the application or user preferences. The other type of graph metric in \cite{schuhmacher2023assignment} is the graph transport-transform (GTT) metric. The GTT metric is based on the transport-transform metric \cite{muller2020metrics}, which is similar to the generalised optimal sub-pattern assignment (GOSPA) metric ($\alpha=2$) for sets \cite{Rahmathullah17}. The GTT metric is unnormalised and establishes correspondences between nodes in the two graphs, allowing nodes to remain unassigned. This metric therefore penalises node attribute errors for assigned nodes, number of missed nodes, number of false nodes and also edge mismatches.

The graph GOSPA metric is another metric for graphs of different sizes that makes assignments between nodes in both graphs, with the possibility of leaving some nodes unassigned \cite{gu24}. The graph GOSPA metric also has penalties for node attribute errors for properly assigned nodes, number of missed nodes, number of false nodes and edge mismatches. An important benefit of the graph GOSPA metric is that the definition of the edge mismatches enables a fast implementation (polynomial time) via the linear programming (LP) relaxation of the metric, which is also a metric. Specifically, in the graph GOSPA metric \cite{gu24}, there is an edge mismatch if, between two pairs of nodes that are assigned, the edge exists in one graph but does not in the other graph. The graph GOSPA metric also includes half-edge costs for edges that connect an assigned node with an unassigned node. 

The contribution of this paper is the definition of a family of graph GOSPA metrics to compute the distance between graphs of different sizes. The family considers undirected, unweighted graphs, with node attributes and without self-loops. The family of metrics is shown to be a principled, mathematically-defined metric and is also based on the calculation of an optimal assignment between nodes of the two graphs, with the possibility of leaving nodes unassigned. The family of metrics also has a clear interpretability penalising node attribute errors for assigned nodes, the number of missed and false nodes and edge mismatches. The family generalises the graph GOSPA metric by allowing a more flexible choice of the edge mismatch cost. 

In particular, the family defines three types of edge-related errors, which can have different costs. These are: number of assigned edge mismatches (between two assigned nodes), number of half-assigned edges (between one assigned node and an unassigned node), and number of unassigned edges (which connect two nodes that are unassigned). As indicated in Table \ref{tab:property compare}, no previous graph metric in the literature considers these three types of edge costs. For instance,  the graph GOSPA metric considers two types of edge errors: number of assigned edge mismatches and number of half-assigned edges, each one being penalised half the cost of an assigned edge mismatch. Also, the above-mentioned GTT metric considers a single type of cost for all edge mismatches (including assigned, half-assigned and unassigned).

\begin{table}
    \centering
    \caption{Properties of graph distances. 'D': implies that it is a metric for different number of nodes, 'F': it is a metric for a fixed number of nodes.}
    \begin{tabular}{c|c|c|c}
    \hline 
        & Node& Types of  & Metric\\
        &   attributes&edge mismatches & property    \\
    \hline
    MCS~\cite{bunke1997relation,bunke1998graph}& \xmark & 1&  D \\
    GED~\cite{sanfeliu1983}&\cmark & 1 & D \\
    GCD~\cite{Bento19}&\cmark & 1 & F  \\
    Graph OSPA \& GTT~\cite{schuhmacher2023assignment}&\cmark & 1& D \\
    Graph GOSPA~\cite{gu24}&\cmark &2 & D \\
    Graph GOSPA family& \cmark& 3& D \\
    \hline
    \end{tabular}
    \label{tab:property compare}
\end{table}

A second contribution of this paper is the development of a tractable approximation of the graph GOSPA metric family that is calculated via LP. Experimental results show the increased flexibility of the metric, compared to graph GOSPA, and the benefits in graph classification via a K-nearest neighbours (KNN) classifier \cite{cover1967nearest}.

The rest of the paper is organised as follows. In Section \ref{sec:Problem-formulation}, we provide the definition of the graph metrics and the background on the GOSPA metric. Section \ref{sec:graph gospa family} defines the proposed graph GOSPA metric family and provides illustrative examples on how the metric works. This section also presents the metric decomposition, including attribute errors for assigned nodes, false and missed errors for unassigned nodes, and errors for assigned, half-assigned and unassigned edges. Section \ref{subsec:approximate computation} presents the computation of the metric family based on a mixed-integer quadratic program (MIQP) and a mixed-integer LP (MILP). It also presents the approximate computations via relaxation of the binary constraints. Section \ref{sec:experimental results} presents the experimental results. The conclusions are drawn in Section \ref{sec:conclusion}.

\section{Problem formulation and background}\label{sec:Problem-formulation}
In this section, we formulate the problem of defining metrics for graphs and also review the GOSPA metric.

\subsection{Metrics for graphs}\label{subsec:metric for graphs}

An undirected graph $G \in \Omega$, where $\Omega$ is the space of undirected graphs without self-loops, can be defined by $G=(V,E)$, a pair containing vertices (or nodes) $V$ and a set of edges $E$. In a vertex set $V=\{x_1,\dots,x_n\}$, the $i$-th vertex is denoted by $x_i \in \mathbb{X}$, where $\mathbb{X}$ is the attribute space of the vertex, with $\mathbb{X}=\mathbb{R}^N$ if the attribute is a real vector. Thus, $V \in \mathcal{F}(\mathbb{X})$, where $\mathcal{F}(\mathbb{X})$ denotes the set of finite subsets of $\mathbb{X}$. The edge set $E$ is defined based on the vertex set where $E(V) \subseteq \mathbb{E}=\{\{x,y\}:x,y\in V, x \neq y\}$. 

To define a metric $d(\cdot,\cdot)$ between two graphs $X=(V_X, E_X)$ and $Y=(V_Y, E_Y)$, the metric properties need to be considered. A distance function $d(\cdot,\cdot): \Omega \times \Omega \rightarrow \mathbb{R}_+$, is a metric on $\Omega$ if it meets the following properties for all $X,Y,Z \in \Omega$ \cite{Apostol_book74}:
\begin{enumerate}
    \item $d(X,Y)=0$ if and only if $X=Y$ (identity);
    \item $d(X,Y)=d(Y,X)$ (symmetry);
    \item $d(X,Y) \leq d(X,Z) + d(Z,Y)$ (triangle inequality).
\end{enumerate}

\subsection{GOSPA metric}\label{subsec:gospa metric}
This section briefly reviews the GOSPA metric, as it is the base for the graph GOSPA family of metrics \cite{Rahmathullah17}. 

The GOSPA metric is a metric between two sets of nodes and is based on node assignments. 
Suppose we have two sets of nodes $V_X=\{x_1,...,x_{n_X}\}$ and $V_Y=\{y_1,...,y_{n_Y}\}$. Let $\gamma$ be an assignment set between $\{1,\dots,n_X\}$ and $\{1,\dots,n_Y\}$, where $\gamma \subseteq \{1,\dots,n_X\} \times \{1,\dots,n_Y\}$. That is, $\gamma$ is a subset of the set of all possible assignments $\Gamma$. Then we denote the set of all possible $\gamma$ as $\Gamma$.

\begin{defn}\label{defn:gospa definition}
Given a scalar $c>0$, a scalar $p$ with $1 \leq p \leq \infty$, and a base metric $d(\cdot,\cdot)$ on the attribute space $\mathbb{X}$, the GOSPA metric ($\alpha =2$) \cite{Rahmathullah17} between $V_X$ and $V_Y$ is 
\begin{flalign}
    d^{\left(c\right)}&\left(V_X,V_Y\right) \nonumber \\
 &=\min_{\gamma \in \Gamma} \left( \sum_{(i,j)\in \gamma} d^p(x_i,y_j)+\frac{c^p}{2}(n_X+n_Y-2|\gamma|) \right)^{1/p} \label{eq:gospa metric}.
\end{flalign}
\end{defn}

In \eqref{eq:gospa metric}, the first term represents the distance between node attributes for assigned nodes to the power of $p$ and the second term represents the cost of unassigned nodes, to the power of $p$. Assigned nodes $x_i$ and $y_{\pi_i}$ contribute $d(x_i,y_{\pi_i})^p$ to the overall error, and any unassigned node in $V_X=\{x_1,\dots,x_{n_X}\}$ or $V_Y=\{y_1,\dots,y_{n_Y}\}$ is penalised with $c^p/2$.

When applied to graphs, the GOSPA metric only computes the differences between node attributes and does not consider the corresponding edges. Thus, the GOSPA metric does not satisfy the identity property and is not a metric for graphs.

\section{Graph GOSPA metric family}\label{sec:graph gospa family}
In this section, we introduce the graph GOSPA metric family in terms of assignment sets
in Section \ref{subsec:graph gospa family}. Section \ref{subsec:illustrative examples} provides illustrative examples of how this metric family works. Section \ref{subsec:graph GOSPA metric family assignment matrices} shows how to write the graph GOSPA metric family in terms of assignment matrics. In Section \ref{subsec: metric decompose}, we show the decomposition of the metric family.

\subsection{Graph GOSPA family in terms of assignment sets}\label{subsec:graph gospa family}
Suppose we have two graphs $X=(V_X,E_X)$ and $Y=(V_Y,E_Y)$ with sets of nodes $V_X=\{x_1,...,x_{n_X}\}$ and $V_Y=\{y_1,...,y_{n_Y}\}$. The corresponding adjacency matrices for the sets of edges with node ordering $(x_1,...,x_{n_X})$ and $ (y_1,...,y_{n_Y})$ are $A_X$ and $A_Y$, respectively. The $(i,j)$ element of matrix $A_X$ is denoted as $A_X(i,j)$, where $A_X(i,j) \in \{0,1\}$. If node $i$ and node $j$ are connected by an edge, $A_X(i,j)=A_X(j,i)=1$, otherwise, $A_X(i,j)=A_X(j,i)=0$. As we consider graphs without self-loops, $A_X(i,i) = 0$ for all $i$.

Similarly to GOSPA, the graph GOSPA metric family makes assignments between nodes, and includes costs for assigned nodes and unassigned nodes. Additionally, it includes edge mismatch costs. Given the assignment set $\gamma$, the sets of unassigned node indices in $X$ and $Y$ are $U_{X}\left(\gamma\right)=\left\{ 1,...,n_{X}\right\} \setminus\left\{ i:\left(i,j\right)\in\gamma\right\}$ and
$U_{Y}\left(\gamma\right)=\left\{ 1,...,n_{Y}\right\} \setminus\left\{ j:\left(i,j\right)\in\gamma\right\} $. Then, we can formally define the graph GOSPA metric family.

\begin{defn}\label{defn:definition of metric family}
For the hyperparameters $c$, $p$, $\epsilon$, $\beta$, $\eta$, such that $c>0$, $1\leq p<\infty$, $\epsilon >0$ and $0\leq\beta\leq \eta \leq 1$, the graph GOSPA metric family between graph $X$ and graph $Y$ is 
    \begin{flalign}
        &d(X,Y)= \nonumber\\ 
        &\min_{\gamma \in \Gamma} \left( \sum_{(i,j)\in \gamma} d^p(x_i,y_j)+\frac{c^p}{2}(|V_X|+|V_Y|-2|\gamma|)+e^p(\gamma) \right) ^{1/p},
    \end{flalign}\label{eq:gospa family defn}
where the total edge mismatch error (to the $p$-th power) is
\begin{flalign}
        &e^p(\gamma)=\frac{\epsilon^p}{2}\sum_{(i_1,j_1)\in \gamma} \sum_{(i_2,j_2)\in \gamma}\left|A_X(i_1,i_2)-A_Y(j_1,j_2)\right| \nonumber \\
        &+\eta \epsilon^p \sum_{(i_1,j_1)\in \gamma} \left[\sum_{i_2 \in U_X(\gamma)}A_X(i_1,i_2)+\sum_{j_2 \in U_Y(\gamma)}A_Y(j_1,j_2)\right] \nonumber \\
        &+\frac{\beta}{2} \epsilon^p \left[\sum_{i_1\in U_X(\gamma)} \right.  \sum_{i_2\in 
        U_X(\gamma)}A_X(i_1,i_2) \nonumber \\ 
        &+ \left. \sum_{j_1\in U_Y(\gamma)}
        \sum_{j_2\in U_Y(\gamma)}A_Y(j_1,j_2)\right] \label{eq:edge cost}  
\end{flalign}
where $\epsilon^p$ is the assigned edge mismatch cost (to the $p$-th power), $\eta \epsilon^p$ is the half-assigned edge cost (to the $p$-th power), $\beta \epsilon^p$ is the unassigned edge cost (to the $p$-th power), and we use $|V_X|$ and $|V_Y|$ to denote number of nodes in graphs $X$ and $Y$, respectively. 
\end{defn}
The identity and symmetry properties of \eqref{eq:gospa family defn} hold directly. The proof of the triangle inequality of $d(\cdot,\cdot)$ is provided in Appendix A in the supplementary material. It should be noted that to ensure that the triangle inequality holds it is required that these different costs have the same exponent $p$ and that  $0\leq\beta\leq \eta \leq 1$.

Analogously to the GOSPA metric in \eqref{eq:gospa metric}, the graph GOSPA metric family in \eqref{eq:gospa family defn} is also an optimisation problem over assignment sets. In the graph GOSPA metric family, we optimise the same function as in \eqref{eq:gospa metric}, including errors in node attributes (localisation errors) and costs for missed and false nodes, but with an additional penalty for edge mismatches $e(\gamma)$. If a node in $X$ is assigned to another node in $Y$, we denote them as assigned nodes.
The edge mismatch cost can be divided into three types:
\begin{itemize}
    \item If nodes $i_1$ and $i_2$ in $X$ are assigned to nodes $j_1$ and $j_2$ in $Y$, and there is an edge between nodes $i_1$ and $i_2$ in $X$ but no edge between $j_1$ and $j_2$ in $Y$ (and vice versa), this leads to an assigned edge mismatch $\epsilon^p$ (the first line in \eqref{eq:edge cost}). This line has a multiplication factor $1/2$ as the sum counts these edge mismatches twice.
    \item If there is an edge between nodes $i_1$ and $i_2$ in $X$, and node $i_1$ in $X$ is assigned to the node $j_1$ in $Y$, and node $i_2$ in $X$ is left unassigned, this leads to a half assigned edge error $\eta\epsilon^p$ (the second line in \eqref{eq:edge cost}). 
    \item If there is an edge in $X$ or $Y$ has both nodes left unassigned, that leads to an unassigned edge error $\beta\epsilon^p$ (the third and fourth lines in \eqref{eq:edge cost}). These lines include a multiplication factor $1/2$, as the sums also count these edge mismatches twice.
\end{itemize}

In the graph GOSPA metric family, $c$, $p$, $\epsilon$, $\eta$, $\beta$ are hyperparameters. By changing $c$, we can control the penalty for unassigned nodes, and sensitivity to node attribute errors. We can also adjust $\epsilon$ to change the sensitivity to the mismatches in edges. Moreover, by changing $\beta$ and $\eta$, we can adapt the weight of the penalties of half-assigned and unassigned edges.

It should be noted that the graph GOSPA metric \cite{gu24} can be recovered by setting $\beta=0$, $\eta=1/2$, meaning that it does not penalise unassigned edge errors, and that half-assigned edges are penalised $1/2$ of an assigned edge mismatch. 
A benefit of the graph GOSPA metric family compared to the graph GOSPA metric is that there is more flexibility to choose how to penalise edge mismatches. However, the graph GOSPA metric family is in general more computationally demanding to calculate, compared to the LP implementation of the graph GOSPA metric \cite{gu24}.

\subsection{Illustrative examples}\label{subsec:illustrative examples}
In this section, we use the examples in Figure \ref{fig:eg1} to demonstrate how the graph GOSPA metric family works. For simplicity, we consider that $p=1$, and there is no localisation error between assigned nodes.

\tikzset{x=1.6 cm,y=1.6 cm, every text node part/.style={align=center}, every node/.style={font=\scriptsize, inner sep=1pt,outer sep=0pt, minimum size=2pt,  draw}}
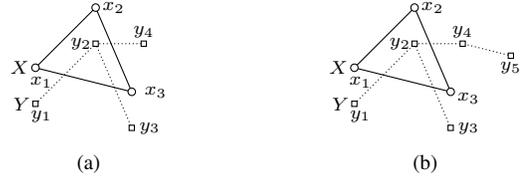
\begin{figure}[t!]
\begin{center}

 \begin{minipage}[t]{0.24\textwidth}
\centering
\begin{tikzpicture}
    \def \len {2.5}; 	\def \lenn {5};\def \del{0.3}; \def \ylim {-0.5};
    
    \node [circle, label =left:$X$](x1) at (1, 0.5){};
    \node [circle](x2) at (1.5,1){};
    \node [circle](x3) at (1.8,0.3){};
    \draw (x1) node[draw=none] at (1.05, 0.38) {$x_1$} --(x2) node[draw=none] at (1.65, 1) {$x_2$};
    \draw (x2)--(x3) node[draw=none] at (2.0, 0.3) {$x_3$};
    \draw (x1)--(x3) node[draw=none] at (1,1) {};
    \begin{scope}[blue]
    \node [rectangle, label =left:$Y$](y1) at (1, 0.5-\del){};
    \node [rectangle](y2) at (1.5,1-\del){};
    \node [rectangle](y3) at (1.8,0.3-\del){};
    \node [rectangle](y4) at (1.9,0.7){};

    \draw [densely dotted](y1) node[draw=none] at (1.05,0.1) {$y_1$} --(y2) node[draw=none] at (1.38,0.7){$y_2$};
    \draw [densely dotted](y2) node[draw=none] at (1.95,0){$y_3$}-- (y3) node[draw=none] at (1.55,0.6){};
    \draw [densely dotted](y4) node[draw=none] at (1.9,0.8){$y_4$} -- (y2) node[draw=none] at (1.55,0.6){};
    \end{scope}

 \end{tikzpicture}
\subcaption{}
\label{fig:eg1a}
\end{minipage}
\begin{minipage}[t]{0.24\textwidth}
\centering
\begin{tikzpicture}
    \def \len {2.5}; 	\def \lenn {5};\def \del{0.2}; \def \ylim {-0.5};

    \def \len {2.5}; 	\def \lenn {5};\def \del{0.3}; \def \ylim {-0.5};

    \node [circle, label =left:$X$](x1) at (1, 0.5){};
    \node [circle](x2) at (1.5,1){};
    \node [circle](x3) at (1.8,0.3){};
    \draw (x1) node[draw=none] at (1.05, 0.38) {$x_1$} --(x2) node[draw=none] at (1.65, 1) {$x_2$};
    \draw (x2)--(x3) node[draw=none] at (1.95, 0.25) {$x_3$};
    \draw (x1)--(x3) node[draw=none] at (1,1) {};
    \begin{scope}[blue]
    \node [rectangle, label =left:$Y$](y1) at (1, 0.5-\del){};
    \node [rectangle](y2) at (1.5,1-\del){};
    \node [rectangle](y3) at (1.8,0.3-\del){};
    \node [rectangle](y4) at (1.9,0.7){};
    \node [rectangle](y5) at (2.3,0.6){};

    \draw [densely dotted](y1) node[draw=none] at (1.05,0.1) {$y_1$} --(y2) node[draw=none] at (1.38,0.7){$y_2$};
    \draw [densely dotted](y2) node[draw=none] at (1.95,0){$y_3$}-- (y3) node[draw=none] at (1.55,0.6){};
    \draw [densely dotted](y4) node[draw=none] at (1.9,0.8){$y_4$} -- (y2) node[draw=none] at (1.55,0.6){};
    \draw [densely dotted](y5) node[draw=none] at (2.3,0.5){$y_5$} -- (y4) node[draw=none] at (1.9,0.8){};
    \end{scope}
 \end{tikzpicture}
  \subcaption{}
\label{fig:eg1b}
 \end{minipage}

\caption{Example to illustrate the node and edge mismatch costs for the same ground truth graph $X$, and two estimated graphs $Y$. (a) Three assigned nodes, one unassigned node, one assigned edge mismatch and one half-assigned edge; (b) three assigned nodes, two unassigned nodes, one assigned edge mismatch, one half-assigned edge and one unassigned edge.}

\label{fig:eg1}
\vspace{-0.5cm}
 \end{center}
 \end{figure}

In Figure \ref{fig:eg1a}, we compare the ground truth graph $X$ with nodes $(x_1,x_2,x_3)$ and the estimated graph $Y$ with nodes $(y_1,y_2,y_3,y_4)$. In this example, we compute the distance for the assignment pairs $\gamma= \{(1,1), (2,2), (3,3)\}$ (we assume this is the optimal assignment), and the node $y_4$ in graph $Y$ is left unassigned. Thus, in this example, the cost of the graph GOSPA metric family is composed of the cost of one unassigned node $c/2$, the cost of an assigned edge mismatch $\epsilon$, and the cost of one half-assigned edge $\eta\epsilon$. Therefore, the cost for the graph GOSPA metric family is $c/2+\epsilon+\eta\epsilon$. On the other hand, the graph GOSPA metric cost is $c/2+\epsilon+\epsilon/2$.

For the example in Figure \ref{fig:eg1b}, the ground truth graph $X$ is the same as the one in Figure \ref{fig:eg1a}, but in the estimated graph $Y$ as a result of the optimisation problem, there is an extra node $y_5$. In this example, there is an extra unassigned node cost and an unassigned edge cost besides the costs in Figure \ref{fig:eg1a}. Thus, the cost of the graph GOSPA metric family is $c+\epsilon+\eta\epsilon+\beta\epsilon$. On the other hand, the graph GOSPA metric only counts the cost of the extra unassigned node, resulting in a total cost of $c+\epsilon+\epsilon/2$.

\subsection{Graph GOSPA family in terms of assignment matrices}\label{subsec:graph GOSPA metric family assignment matrices}
In this section, we show that the graph GOSPA metric family can be written in terms of binary assignment matrices instead of assignment sets. We first introduce some additional notation.

An assignment set $\gamma$ can be equivalently written as a binary assignment matrix 
$W \in \mathcal{W}_{X,Y}$, which satisfies the following properties \cite{angel20}:
\begin{flalign}
\sum_{i=1}^{n_{X}+1}W(i,j) &=1,\ j=1,\ldots,n_{Y}\label{eq:binary_constraint1}\\
\sum_{j=1}^{n_{Y}+1}W(i,j) &=1,\ i=1,\ldots,n_{X}\label{eq:binary_constraint2}\\
W(n_{X}+1,&n_{Y}+1) =0,\label{eq:binary_constraint3}\\
&W(i,j)  \in\{0,1\} \quad \forall\ i,j,\label{eq:binary_constraint4}
\end{flalign}
where the $W(i,j)$ is the element in row $i$ and column $j$ of matrix $W$. If $x_i$ is assigned to $y_j$, $W(i,j)=1$. If $x_i$ is unassigned, $W(i,n_Y +1 )=1$. Similarly, $W(n_X+1,j)=1$ if $y_j$ remains unassigned. Note that these properties of the assignment matrices encode the same properties of the assignment sets explained before Definition \ref{defn:gospa definition}.

We define a column vector with all ones and length $n_X$ as $1_{n_X}$. Similarly, a column vector with all ones and length $n_Y$ is defined as $1_{n_Y}$. We also use the entry-wise 1-norm for matrices. This is defined as follows for a matrix $A$ of size $n \times m$
\begin{equation}
    \left\Vert A\right\Vert =\sum_{i=1}^{n}\sum_{j=1}^{m}|A(i,j)|.
\end{equation}
Now, we can introduce Lemma \ref{lem:metric family mat}.
\begin{lem}\label{lem:metric family mat}
For the hyperparameters $c$, $p$, $\epsilon$, $\beta$, $\eta$, such that  $c>0$, $1\leq p<\infty$, $\epsilon >0$ and $0 \leq \beta\leq \eta \leq 1$, the graph GOSPA metric family $d_p^{(c,\epsilon)}(\cdot,\cdot)$ in \eqref{eq:gospa family defn} can be written as 
\begin{flalign}
d(X,Y)
&=\min_{\substack{W\in\mathcal{W}_{X,Y}}}\left(\mathrm{tr}\big[D_{X,Y}^{\top}W\big] + e_{X,Y}^{p}(W)\right)^{1/p}\label{eq:simplified_graph_metric_family}
\end{flalign}
where $D_{X,Y}$ is an $(n_X+1) \times (n_Y+1)$ matrix that contains the localisation errors $d^p(x_i,y_j)$ and the penalties for unassigned nodes, such that its $(i,j)$ element is:
\begin{flalign}\label{eq:Dxy}
D_{X,Y}(i,j) =\begin{cases}
d^{p}\left(x_{i},y_{j}\right)\ & i\leq n_{X},j\leq n_{Y},\\
\frac{c^{p}}{2} & i=n_{X}+1,j\leq n_{Y},\\
\frac{c^{p}}{2} & i\leq n_{X},j=n_{Y}+1,\\
0 & i=n_{X}+1,j=n_{Y}+1,
\end{cases}
\end{flalign}
and
\begin{flalign}
e^{p}_{X,Y}&(W) 
 =\frac{\epsilon^{p}}{2} \left\Vert A_{X}W_{1:n_{X},1:n_{Y}}-W_{1:n_{X},1:n_{Y}}A_{Y} \right\Vert\nonumber \\
 & +\frac{\beta}{2}\epsilon^{p}\left(W_{1:n_{X},n_{Y}+1}^{\top}A_{X}W_{1:n_{X},n_{Y}+1} \right. \nonumber \\
 &\left.+W_{n_{X}+1,1:n_{Y}}A_{Y}W_{n_{X}+1,1:n_{Y}}^{\top}\right)\nonumber \\
 & +\left(\eta-\frac{1}{2}\right)\epsilon^{p}\left[W_{1:n_{X},n_{Y}+1}^{\top}A_{X}\left(1_{n_{X}}-W_{1:n_{X},n_{Y}+1}\right) \right. \nonumber \\
 &+\left.\left(1_{n_{Y}}-W_{n_{X}+1,1:n_{Y}}\right)A_{Y}W_{n_{X}+1,1:n_{Y}}^{\top}\right],\label{eq:edge_cost_family}
\end{flalign}
and matrix $W_{1:n_X,1:n_Y}$ of size $n_X \times n_Y$ is formed by the first $n_X$ rows and $n_Y$ columns of $W$. Similarly, $W_{n_{X}+1,1:n_{Y}}$ is a vector of size $1 \times n_Y$ containing the first $n_Y$ elements in row $n_X+1$ of $W$. Finally, $W_{1:n_{X},n_{Y}+1}$ is a vector of size $n_X \times 1$ containing the first $n_X$ elements of column $n_Y+1$ of $W$.
\end{lem}
The first line in \eqref{eq:edge_cost_family} corresponds to the edge mismatch cost in the graph GOSPA metric \cite{gu24}. This includes the assigned edge penalties $\epsilon$ and half-assigned edge penalties (with a cost $\epsilon^p/2$). In this line, matrices $A_{X}W_{1:n_{X},1:n_{Y}}$ and $W_{1:n_{X},1:n_{Y}}A_{Y}$ are of size $n_{X} \times n_{Y}$, and the entry-wise 1-norm outputs a scalar.

The matrix multiplications in the second line in \eqref{eq:edge_cost_family} result in a scalar that counts twice the number of unassigned edges in $A_X$. Therefore, this line is multiplied by $1/2$ to recover the number of unassigned edges, which is then multiplied by $\beta\epsilon^p$. The same argument holds for the third line but for $A_Y$. The matrix multiplications in the fourth line in \eqref{eq:edge_cost_family} provide a scalar that counts the number of assigned-unassigned edges in $A_X$. As the first line has already counted these edge penalties with a cost of $\epsilon^p/2$ (see \cite[Lemma 1]{gu24}), in its multiplication factor we subtract $1/2$ from $\eta$, such that the total number of half-assigned edges is multiplied by $\eta \epsilon^p$. The same argument holds for the fifth row in \eqref{eq:edge_cost_family} but for $A_Y$.

\subsection{Metric decomposition}\label{subsec: metric decompose}

In this section, we explain the decomposition of the family of metrics into its different types of costs: node attribute errors for properly assigned nodes, number of missed nodes, number of false nodes, and the different types of edge mismatches: assigned edge mismatches, half-assigned edge mismatches and unassigned edge mismatches.

Without loss of generality to define the missed and false nodes, we consider that $X$ is the ground truth graph and $Y$ is an estimated graph. If $i \leq n_X, j \leq n_Y$, which we denote as $(i,j)\in S_1$, and $D_{X,Y}$ represents the localisation cost for assigned nodes. If $i \leq n_X$ and $j=n_Y+1$, which we denote as $(i,j) \in S_2$, $D_{X,Y}(i,j)$ represents a missed node cost. If $i=n_X+1$ and $j \leq n_Y$, which we denote as $(i,j)\in S_3$, $D_{X,Y}(i,j)$ represents a false node cost. We can then write the graph GOSPA metric family as below:
\begin{flalign}
    d(X,Y) 
    &=\min_{\substack{W\in\mathcal{W}_{X,Y}}} \bigg( \mathrm{l}(X,Y,W)^p  
    + \mathrm{m}(X,Y,W)^p \nonumber \\
    &+ \mathrm{f}(X,Y,W)^p  
      + e_{X,Y}^p(W) \bigg)^{1/p} 
\end{flalign}
where 
\begin{flalign}
    \mathrm{l}(X,Y,W)^p &= \sum_{(i,j)\in S_1} D_{X,Y}(i,j) W(i,j) \nonumber \\
    \mathrm{m}(X,Y,W)^p &= \frac{c^p}{2} \sum_{(i,j)\in S_2} W(i,j) \nonumber \\
    \mathrm{f}(X,Y,W)^p &=  \frac{c^p}{2} \sum_{(i,j)\in S_3} W(i,j),
\end{flalign}
and they represent the costs (to the $p$-th power) for assigned, missed and false nodes with the assignment $W$. The edge cost $e_{X,Y}^{p}(W)$ is defined in \eqref{eq:edge_cost_family}, which contains the assigned edge costs, half-assigned edge costs and unassigned edge costs. Thus, the graph GOSPA metric family cost admits this decomposition, which provides the interpretability of the family.

\section{Metric family computation}\label{subsec:approximate computation}
In this section, we show that the graph GOSPA metric family can be computed by solving an MIQP problem (Section \ref{subsubsec:qp fomulation}). We also show that the MIQP problem can be linearised and reformulated as an MILP problem (Section \ref{subsubsec:linearisation of quadratic prog}). Finally, Section \ref{subsubsec:relaxation of lp} presents the relaxed versions, which provide approximate solutions, and can be computed via LP and QP. 
\subsection{MIQP formulation}\label{subsubsec:qp fomulation}
The graph GOSPA metric family in \eqref{eq:simplified_graph_metric_family} can be written as:
\begin{flalign}
    \underset{\substack{W\in \mathcal{W}_{X,Y}}}{\arg\min}\left(\mathrm{tr}\big[D_{X,Y}^{\top}W\big]+e_{X,Y}^p(W)\right)^{1/p}\label{eq:general_IP}
\end{flalign}
where
\begin{flalign}
    e^p_{X,Y}(W) =\frac{\epsilon^{p}}{2} \left\Vert A_{X}W_{1:n_{X},1:n_{Y}}-W_{1:n_{X},1:n_{Y}}A_{Y} \right\Vert + \epsilon^p \hat{g}\label{eq:edge_cost_combined}
\end{flalign} 
where
\begin{flalign}
 \hat{g}=&\left(\frac{\beta}{2}-\eta+\frac{1}{2}\right)W_{1:n_{X},n_{Y}+1}^{\top}A_{X}W_{1:n_{X},n_{Y}+1} \nonumber \\
 & +\left(\frac{\beta}{2}-\eta+\frac{1}{2}\right)W_{n_{X}+1,1:n_{Y}}A_{Y}W_{n_{X}+1,1:n_{Y}}^{\top} \nonumber \\
 & +\left(\eta-\frac{1}{2}\right)W_{1:n_{X},n_{Y}+1}^{\top}A_{X}1_{n_{X}} \nonumber \\
 & +\left(\eta-\frac{1}{2}\right)1^{\top}_{n_{Y}}A_{Y}W_{n_{X}+1,1:n_{Y}}^{\top}. 
\end{flalign}

As in \cite{angel20, bento2016metric}, we linearise the absolute value in \eqref{eq:edge_cost_combined}, by adding an extra variable $\hat{e} \in \mathbb{R}$, such that the optimisation problem becomes
\begin{flalign}
    \underset{\substack{W\in \mathcal{W}_{X,Y}},\hat{e}}{\arg\min}\left(\mathrm{tr}\big[D_{X,Y}^{\top}W\big]+ \frac{\epsilon^p}{2}\hat{e}+\epsilon^p\hat{g}\right)^{1/p}\label{eq:optimisation problem}
\end{flalign}
where $\hat{e}$ has the following constraint
\begin{flalign}
    \hat{e} \geq \sum_{i=1}^{n_{X}}\sum_{j=1}^{n_{Y}}\left|\sum_{k=1}^{n_{X}}A_{X}(i,k)W(k,j) -\sum_{k=1}^{n_{Y}}W(i,k)A_{Y}(k,j)\right|. \label{eq:e_constraints}
\end{flalign}
The constraint in \eqref{eq:e_constraints} can be linearised by introducing an additional matrix $H$, where its $(i,j)$ element is $H(i,j) \in \mathbb{R}$ \cite{gu24}, such that
\begin{flalign}
    \hat{e} &\geq \sum_{i=1}^{n_{X}}\sum_{j=1}^{n_{Y}} H(i,j), \label{eq:e>H}\\
    H(i,j) &\geq \sum_{k=1}^{n_{X}}A_{X}(i,k)W(k,j) -\sum_{k=1}^{n_{Y}}W(i,k)A_{Y}(k,j), \label{eq:h>abs}\\
    H(i,j) &\geq \sum_{k=1}^{n_{Y}}W(i,k)A_{Y}(k,j) -\sum_{k=1}^{n_{X}}A_{X}(i,k)W(k,j). \label{eq:h>-abs}
\end{flalign}

Now, we write \eqref{eq:optimisation problem} in quadratic form. 
We first define the column vectors $l=[(A_X1_{n_X})^{\top},1_{n_Y}^{\top}A_Y]^{\top}$ and $x=[W_{1:n_X,n_Y+1}^{\top}, W_{n_X+1,1:n_Y}]^{\top}$. We also denote the $i$-th component of vector $x$ as $x(i)$. Note that $x$ is a rewriting of some elements in the optimisation variable $W$. Then this yields
\begin{flalign}
    &\underset{\substack{W\in \mathcal{W}_{X,Y}},\hat{e},H}{\arg\min} \nonumber \\
    &\left( \left(\frac{\beta}{2}-\eta+\frac{1}{2}\right)\epsilon^p\sum_{i=1}^{n_X+n_Y} \sum_{j=1}^{n_X+n_Y} Q(i,j)x(i)x(j) \right.\nonumber \\
    &\left.+ \left(\eta-\frac{1}{2}\right)\epsilon^p\sum_{i=1}^{n_X+n_Y} l(i)x(i) + \mathrm{tr}\big[D_{X,Y}^{\top}W\big]+ \frac{\epsilon^p}{2}\hat{e}\right)^{1/p} \label{eq:quad_prog}
\end{flalign}
where
$Q=\begin{bmatrix}
    A_X & 0 \\
    0 & A_Y
\end{bmatrix}
$, and we should notice that $Q$ is a square matrix of size $(n_X+n_Y) \times (n_X+n_Y)$.

Thus, \eqref{eq:quad_prog} with constraints \eqref{eq:binary_constraint1}, \eqref{eq:binary_constraint2}, \eqref{eq:binary_constraint3}, \eqref{eq:binary_constraint4}, \eqref{eq:e>H}, \eqref{eq:h>abs} and \eqref{eq:h>-abs} is an MIQP problem, with $W(i,j) \in \{0,1\}$.

\subsection{MILP formulation}\label{subsubsec:linearisation of quadratic prog}
As the quadratic term $Q$ in \eqref{eq:quad_prog} is composed of two adjacency matrices, it can be non-convex, and the computational time for finding the integer solution of the non-convex quadratic programming problem is NP-hard \cite{pardalos1991quadratic}. In this case, it is a common procedure to add auxiliary variables to convert the MIQP problem to an MILP problem to speed up computation \cite{furini2019theoretical,Adams04,glover1975improved}. In particular, we convert the MIQP problem into an MILP problem with Glover linear formulation \cite{glover1975improved} by following the procedure in \cite[Sec. 2.2]{furini2019theoretical}. The details of the formulation of the MILP problem are provided in the following.

To write the MIQP problem \eqref{eq:quad_prog} in linear form with the Glover linearisation method \cite{glover1975improved}, we first introduce an auxiliary variable $q \in \mathbb{R}^{(n_X+n_Y)}$ with the following constraints:
\begin{flalign}
    &q(i) \leq Q(i)^+x(i) \label{eq:qp_constraint1}&\\
    &q(i) \geq Q(i)^-x(i) \label{eq:qp_constraint2}&\\
    &q(i) \leq \sum_{j=1}^{n_X+n_Y}Q(i,j)x(j)-Q(i)^-(1-x(i))&  \label{eq:qp_constraint3} \\
    &q(i) \geq \sum_{j=1}^{n_X+n_Y}Q(i,j)x(j)-Q(i)^+(1-x(i))& \label{eq:qp_constraint4}
\end{flalign}
where $i=1,\ldots,n_{X}+n_Y$ and $Q^+$ and $Q^-$ are
\begin{flalign}
Q^-(i)=\sum_{j=1}^{n_X+n_Y} \mathrm{min} \{0,Q(i,j)\} \\
Q^+(i)=\sum_{j=1}^{n_X+n_Y} \mathrm{max} \{0,Q(i,j)\}.
\end{flalign}

Then \eqref{eq:quad_prog} can be written as
\begin{flalign}
    &\underset{\substack{W\in \mathcal{W}_{X,Y}},\hat{e},H,q}{\arg\min} \left(\left(\frac{\beta}{2}-\eta+\frac{1}{2}\right)\epsilon^p\sum_{i=1}^{n_X+n_Y} q(i) \right.\nonumber  \\
    &\left.+\left(\eta-\frac{1}{2}\right)\epsilon^p\sum_{i=1}^{n_X+n_Y} l(i)x(i)+\mathrm{tr}\big[D_{X,Y}^{\top}W\big]+ \frac{\epsilon^p}{2}\hat{e}\right)^{1/p}  \label{eq:linear qp}
\end{flalign}
with the constraints \eqref{eq:binary_constraint1}-\eqref{eq:binary_constraint4}, \eqref{eq:e_constraints}-\eqref{eq:h>-abs} and \eqref{eq:qp_constraint1}-\eqref{eq:qp_constraint4}. Equation \eqref{eq:linear qp} shows that \eqref{eq:optimisation problem} can be solved using MILP.

\subsection{Relaxed LP and QP formulations}\label{subsubsec:relaxation of lp}

The MIQP problem \eqref{eq:quad_prog} and MILP problem \eqref{eq:linear qp} are expensive to compute. With the relaxation of the binary constraints of $W$ in \eqref{eq:binary_constraint4} such that
\begin{align}
    W(i,j) \geq 0, \forall i,j \label{eq:relaxed_constraint}
\end{align}
and keeping the rest of the constraints of \eqref{eq:quad_prog} or  \eqref{eq:linear qp}, we can find lower bounds of the metric family that are faster to compute. In addition, if the solutions have only 0 or 1 values, then they coincide with the non-relaxed graph GOSPA metric family. The relaxation of the MIQP problem yields a (relaxed) QP problem, though this is still complicated to compute due to the non-convexity. The relaxation of the MILP problem yields an LP, which can be calculated in polynomial time. For both the LP and QP relaxations of the graph GOSPA metric family, it is required that $\eta \geq 1/2$ to satisfy the non-negativity property. In addition, the relaxed versions do not meet the triangle inequality in general, as has been checked via simulations. Nevertheless, the relaxed LP and QP versions are metrics for the special case that $\beta = 0$ and $\eta = \frac{1}{2}$, as this corresponds to the LP relaxation of the graph GOSPA metric given in \cite{gu24}. Finally, Table \ref{tab:Optimization_problems} summarises the different types of optimisation problems to compute the metric family. 

\begin{center}
\begin{table}
\caption{Optimisation problems to compute the graph GOSPA metric family}\label{tab:Optimization_problems}

\centering{}%
\begin{tabular}{c|cc}
\hline 
Optimisation & Cost function & Constraints\tabularnewline
\hline 
General & \eqref{eq:general_IP} & \eqref{eq:binary_constraint1}-\eqref{eq:binary_constraint4} \tabularnewline
MIQP &  \eqref{eq:quad_prog} & \eqref{eq:binary_constraint1}-\eqref{eq:binary_constraint4}, \eqref{eq:e>H}-\eqref{eq:h>-abs} \tabularnewline
(Relaxed) QP & \eqref{eq:quad_prog} & \eqref{eq:binary_constraint1}-\eqref{eq:binary_constraint3}, \eqref{eq:e>H}-\eqref{eq:h>-abs}, \eqref{eq:relaxed_constraint} \tabularnewline
MILP & \eqref{eq:linear qp} & \eqref{eq:binary_constraint1}-\eqref{eq:binary_constraint4}, \eqref{eq:e>H}-\eqref{eq:h>-abs}, \eqref{eq:qp_constraint1}-\eqref{eq:qp_constraint4} \tabularnewline
(Relaxed) LP & \eqref{eq:linear qp} & \eqref{eq:binary_constraint1}-\eqref{eq:binary_constraint3}, \eqref{eq:e>H}-\eqref{eq:h>-abs}, \eqref{eq:qp_constraint1}-\eqref{eq:qp_constraint4}, \eqref{eq:relaxed_constraint} \tabularnewline
\hline 
\end{tabular}
\end{table}
\par\end{center}

\section{Experimental results}\label{sec:experimental results}
This section demonstrates the properties of the graph GOSPA metric family\footnote{Python code is available at https://github.com/JinhaoGu/Graph\_GOSPA\_family.} by comparing it with these metrics/distances: the MCS distance \cite[Thm. 1]{bunke1997relation}, the GED \cite{fischer2015approximation} and the GCD \cite{Moharrer20}. We illustrate the increased flexibility of the graph GOSPA metric family with respect to the graph GOSPA metric and the benefits in their interpretability with respect to the other considered distances. 

Section \ref{subsec:simulations} analyses the metrics via simulated graphs and also shows the decomposition of the graph GOSPA metric family into its different components: node attribute error, false node, missed node, and the different types of edge penalties. Section \ref{subsec:evaluate hyperparameters} examines the effects of hyperparameters $\beta$ and $\eta$ on the resulting error. In Section \ref{subsection: computation time}, we analyse the computation time for the graph GOSPA metric family when applied to different sizes of graphs. In Section \ref{subsec: molecule experiment}, we apply the graph GOSPA metric family to real-world datasets for classification purposes and show the benefits compared to other graph metrics. For all the experiments in this section, we use the relaxed LP for the graph GOSPA metric family as the default setting, unless otherwise specified.

\subsection{Evaluation of different metrics}\label{subsec:simulations}
In this section, we compare different metrics. For the graph GOSPA metric family, we use the relaxed LP version, see Section~ \ref{subsec:approximate computation} and Table \ref{tab:Optimization_problems}. The solver for the LP is the interior-point method \cite{luenberger15}. The hyperparameters are set to $c=3$, $p=1$, $\epsilon=1$, $\beta=0.3$, $\eta=0.7$. We have also implemented the (relaxed) QP version, see Section~ \ref{subsec:approximate computation} and Table \ref{tab:Optimization_problems}, using the IBM CPLEX optimiser, yielding basically the same results as the LP implementation, but with a higher computational complexity, so the results are omitted for the sake of clarity.

The hyperparameters in GCD are set to $p=1$, $\lambda=1$ and dummy nodes, with zero attributes and no connection to other nodes, are added to the graph with fewer nodes as in \cite{Bento19}. The MCS distance is computed using the functions in the RDkit library \cite{rdkit} without using any node attributes. For the GED, we use the function in the library NetworkX \cite{networkx11}, and the node match in GED is based on the distance between two nodes, if the distance is less than 3, they will be treated as matched nodes. Moreover, the insertion and deletion costs for nodes are both set to 1.

We use randomly generated graphs from the Erdos-Renyi model \cite{erdds1959random}, with 14 nodes and the probability of an edge existing between two nodes is 0.6. The graphs contain 2-D node attributes in each node. These graphs are generated using the Python library NetworkX \cite{networkx11}. The randomly generated graph $X$ is used as the ground truth and the modification of it is used as the target graph $Y$, which varies across different cases. We analyse four cases in which the graph $Y$ is obtained by (a) adding noise to node attributes, (b) randomly adding or (c) removing edges or (d) removing nodes. By comparing $X$ and $Y$ in these four different cases, we demonstrate the characteristics of the algorithm when there are changes in the structure of graphs.

\begin{figure}[t]
\begin{subfigure}{.24\textwidth}
  \centering
  \includegraphics[width=.999\linewidth]{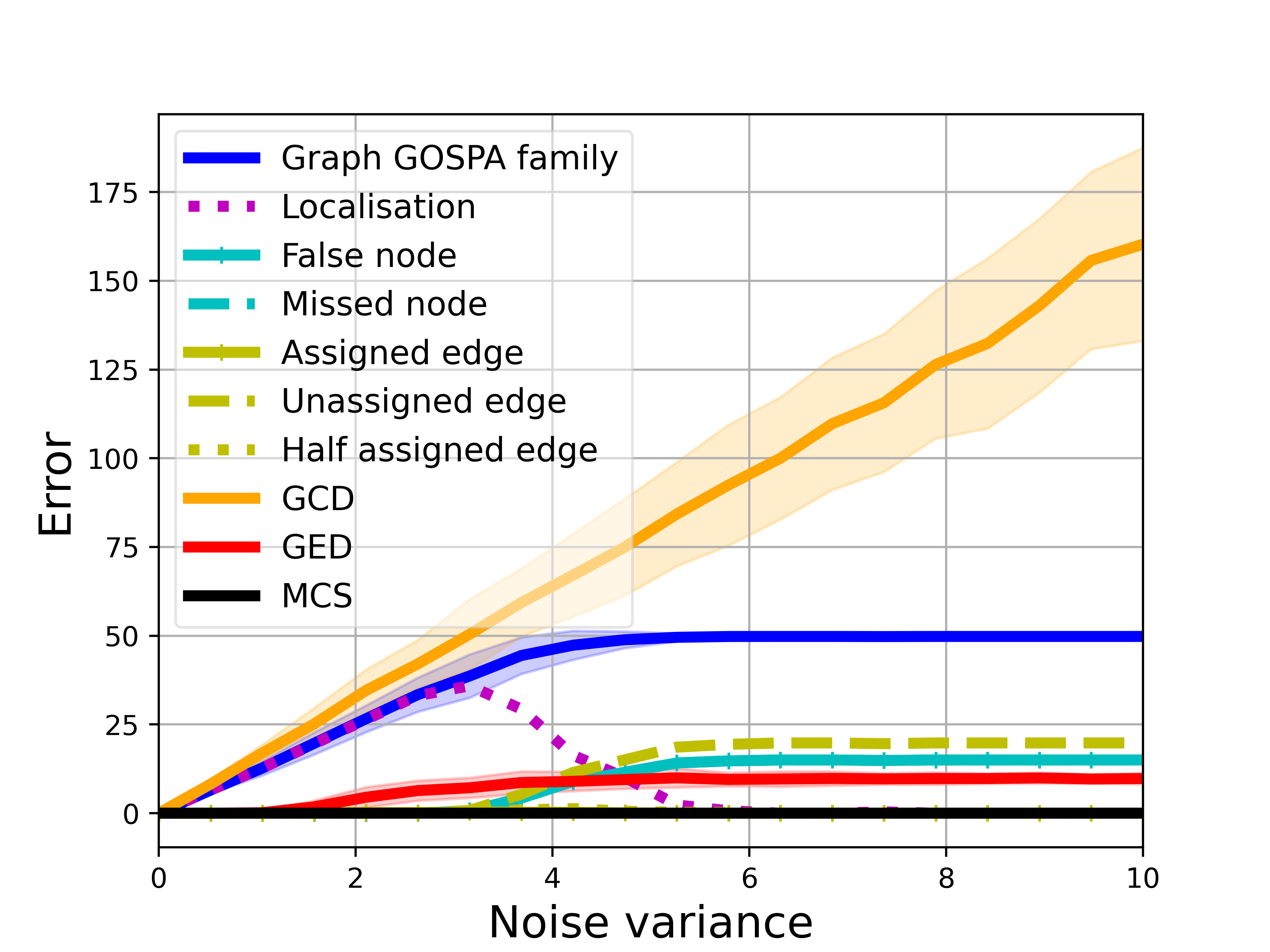}
  \caption{Random node attribute noise.}
  \label{fig:exp_a}
\end{subfigure}
\begin{subfigure}{.24\textwidth}
  \centering
  \includegraphics[width=.999\linewidth]{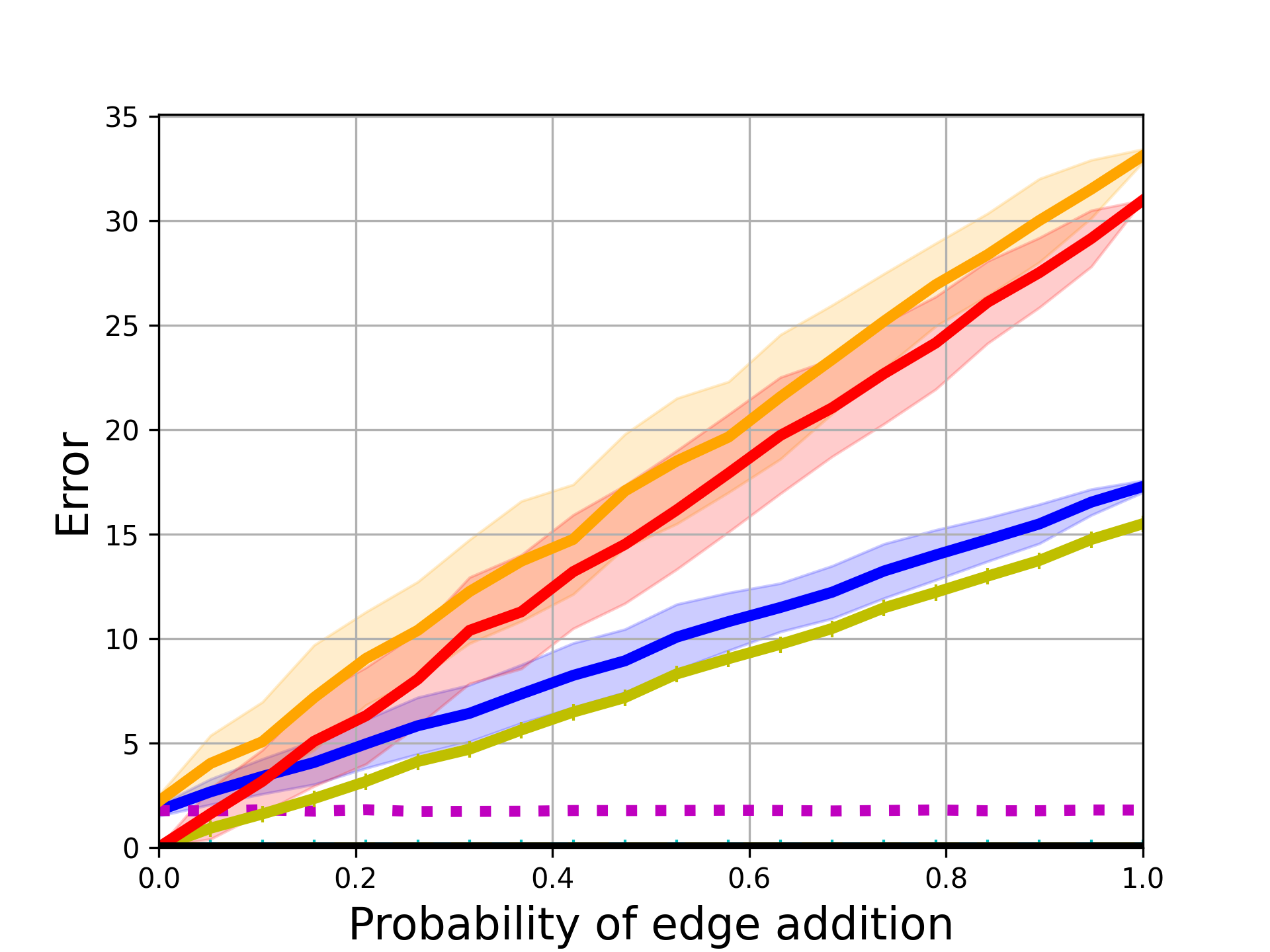}
  \caption{Random edge addition.}
  \label{fig:exp_b}
\end{subfigure}
\begin{subfigure}{.24\textwidth}
  \centering
  \includegraphics[width=.999\linewidth]{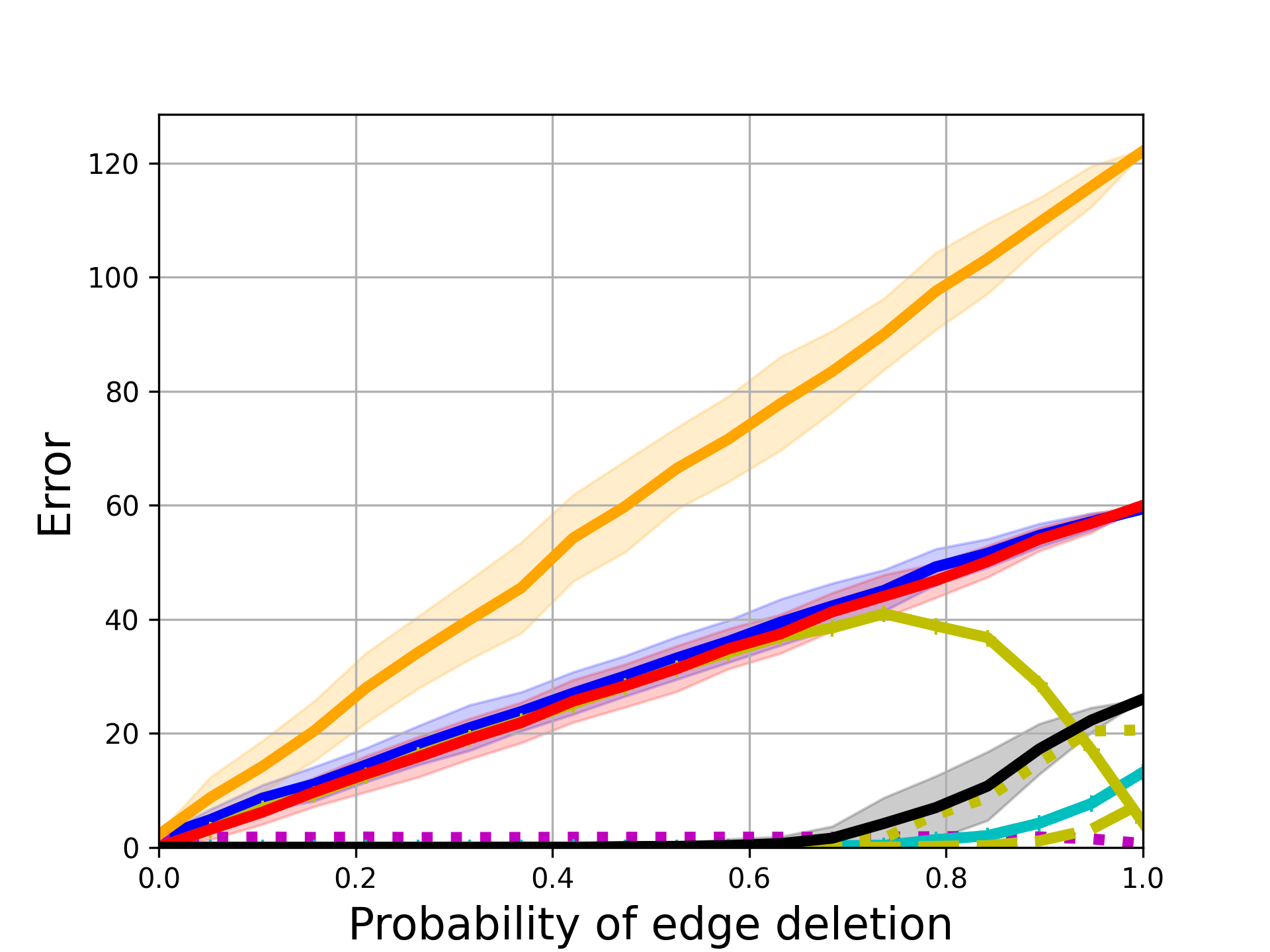}
  \caption{Random edge deletion.}
  \label{fig:exp_c}
\end{subfigure}
\begin{subfigure}{.24\textwidth}
  \centering
  \includegraphics[width=.999\linewidth]{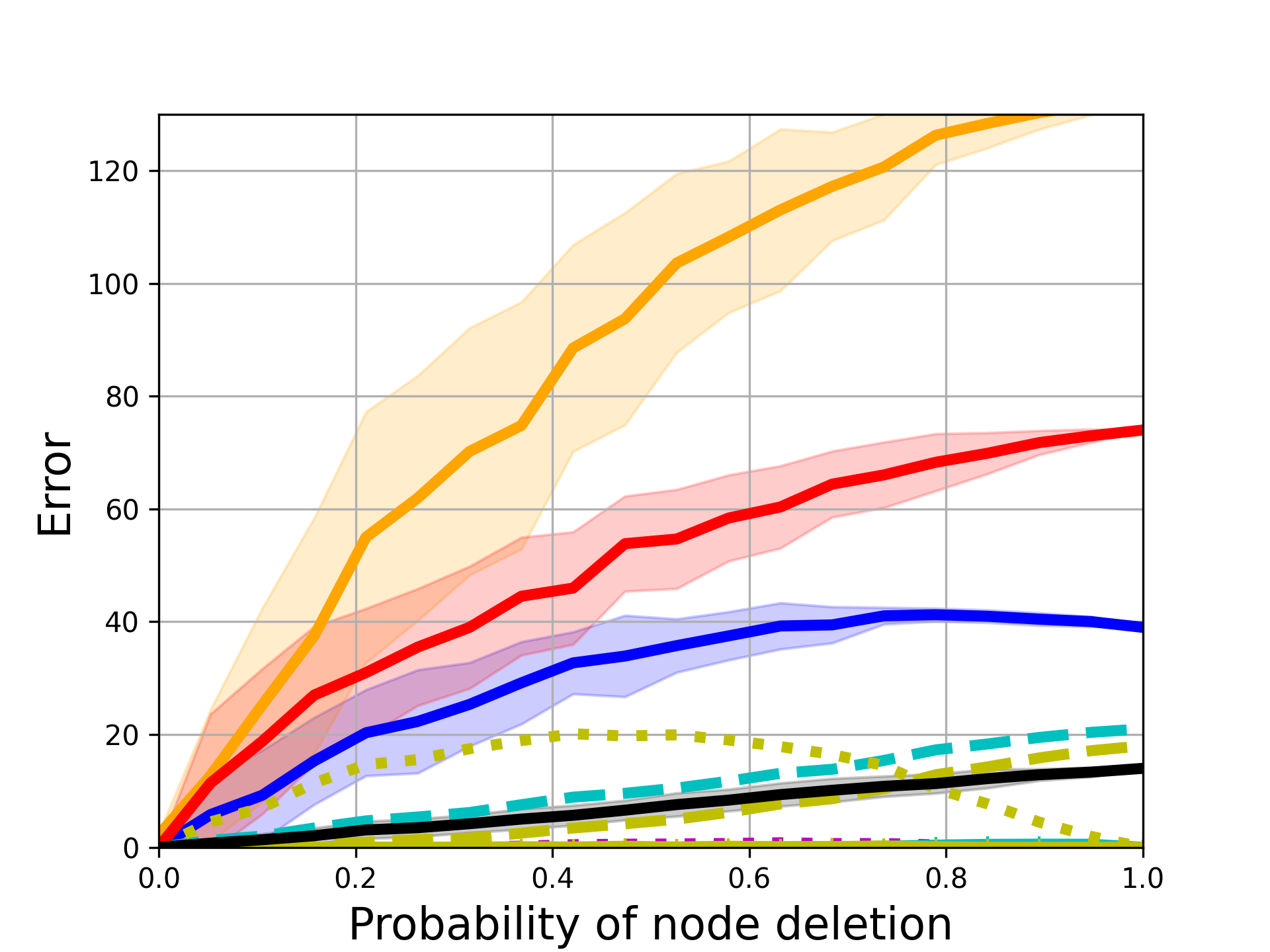}
  \caption{Random node removal.}
  \label{fig:exp_d}
\end{subfigure}
\caption{Plots of average graph GOSPA metric family errors (with the standard deviation shown as a shared area), GCD, GED and MCS, comparing the ground truth graph with (a) graphs with random Gaussian noise with increasing noise variance in node attributes; (b) graphs with random edge addition in adjacency matrix and small Gaussian noise in node attributes; (c) graphs with random edge removal in adjacency matrix and small Gaussian noise in node attributes; (d) graphs with different sizes.}
\label{fig:exp}
\end{figure}

In all four cases, we compute the average distances with 100 Monte Carlo runs, and the standard deviation is shown by the shaded area in each plot. In case (a), we obtain graph $Y$ by adding independent zero-mean Gaussian noise with $\sigma^2 I_2$ to the node attributes in graph $X$. Figure \ref{fig:exp_a} shows the changes of average errors with increasing $\sigma^2$. It can be observed in Figure \ref{fig:exp_a} that the GCD error keeps increasing as the noise level in node attributes increases, due to the growth of localisation errors. In the graph GOSPA family, with the high noise level in node attributes in graph $Y$, the localisation errors are too large to assign two nodes to each other, and the errors of missed nodes and false nodes increase. It is worth noting that the curve for unassigned edges is roughly the same as the curve for unassigned nodes. Since there is no difference between the edges of graph $X$ and graph $Y$, the error between assigned edges is 0. In addition, for increasing noise variance, the standard deviation in the graph GOSPA family becomes smaller since nodes tend to be unassigned and the error does not change across Monte Carlo runs.
The curve for GED errors also has a saturation region, the saturation value depends on the number of nodes in the corresponding graphs. In this case, the MCS distance returns error 0 for all values of noise variance in the node attributes. The reason for this is that the MCS distance does not take into account the node attributes so it is insensitive to the added noise. This is not the desired behaviour, as it is desirable that the distances increase with increasing errors in the node attributes.

In case (b), the graph $Y$ is obtained by adding edges to the adjacency matrix of graph $X$ with the probability $q$. The average errors for different $q$ are given in Figure \ref{fig:exp_b}. In Figure \ref{fig:exp_b}, since all nodes are assigned, there are no unassigned node errors, and we see the errors increase linearly with more edges added. In this case, the MCS distance is also zero. This is because the maximum common subgraph contains all nodes in both graphs when we add more edges. This is not a desired behaviour as it is desired that the distances increase with more edge mismatches.

The graph $Y$ in case (c) is generated by randomly deleting edges in the adjacency matrix of graph $X$ with probability $q$. In Figure \ref{fig:exp_c}, we see that for both GCD and GED, the errors are roughly increasing linearly. For the graph GOSPA family, with more edges being deleted, there are fewer assigned edges, and the assigned edge error decreases until it reaches zero. The error of unassigned edges and false nodes increases, but saturates when almost all edges are deleted.

In case (d), we randomly remove the nodes, with a certain probability, to draw graph $Y$. In Figure \ref{fig:exp_d}, when the probability of node deletion increases, the number of unassigned nodes in the graph GOSPA family grows. The number of half-assigned edges increases first and then drops to zero because the number of assigned nodes decreases. Meanwhile, the number of unassigned edges rises, and the total error of the graph GOSPA family slightly decreases, which depends on the choice of hyperparameters. For other distances, GCD, GED, and MCS distances increase as more nodes are deleted. For GCD, the error has the steepest growth rate of all distances.

\begin{figure}[t]
\begin{subfigure}{.24\textwidth}
  \centering
  \includegraphics[width=.999\linewidth]{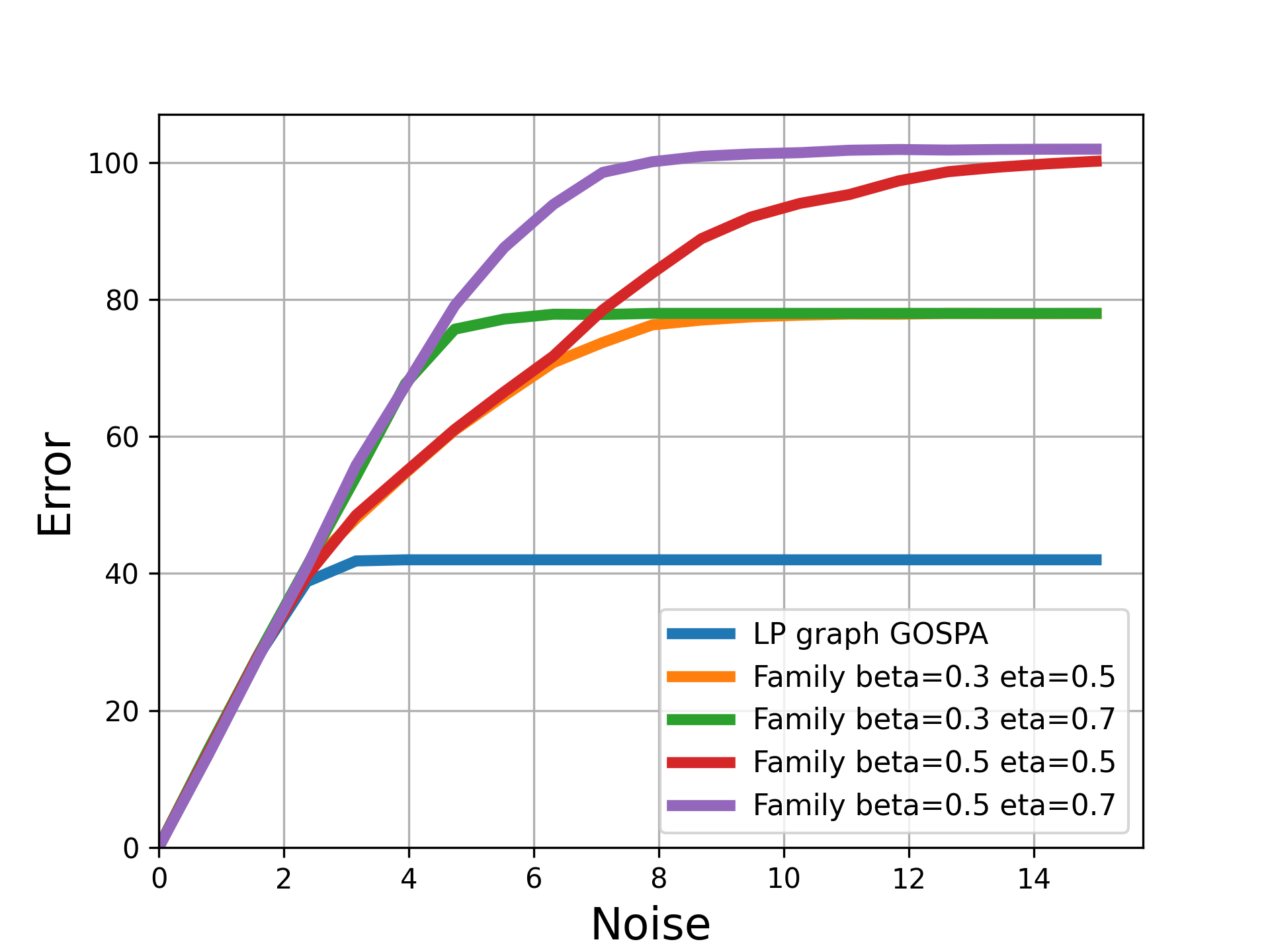}
  \caption{Random node attribute noise.}
  \label{fig:exp2_a}
\end{subfigure}
\begin{subfigure}{.24\textwidth}
  \centering
  \includegraphics[width=.999\linewidth]{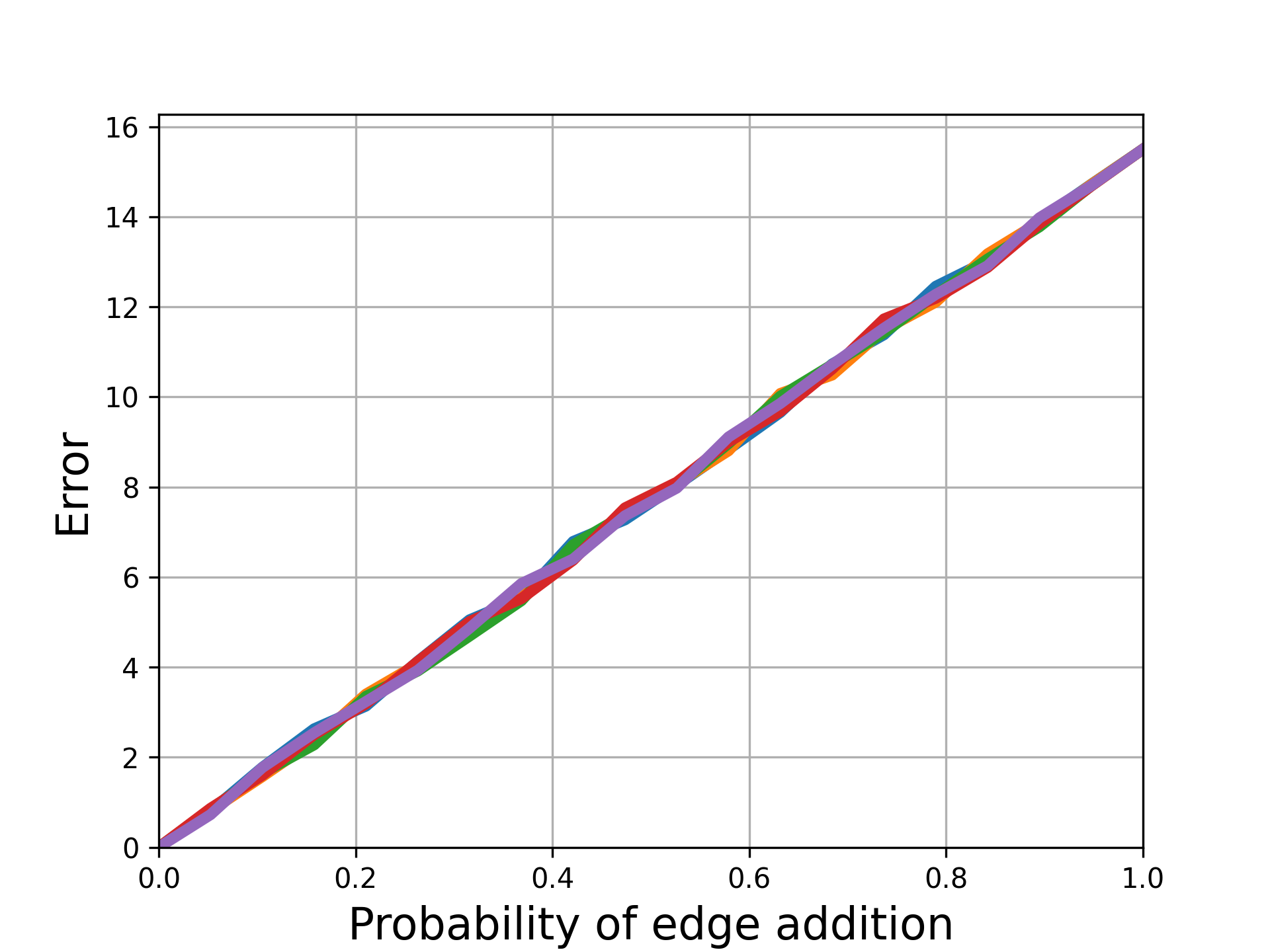}
  \caption{Random edge addition.}
  \label{fig:exp2_b}
\end{subfigure}
\begin{subfigure}{.24\textwidth}
  \centering
  \includegraphics[width=.999\linewidth]{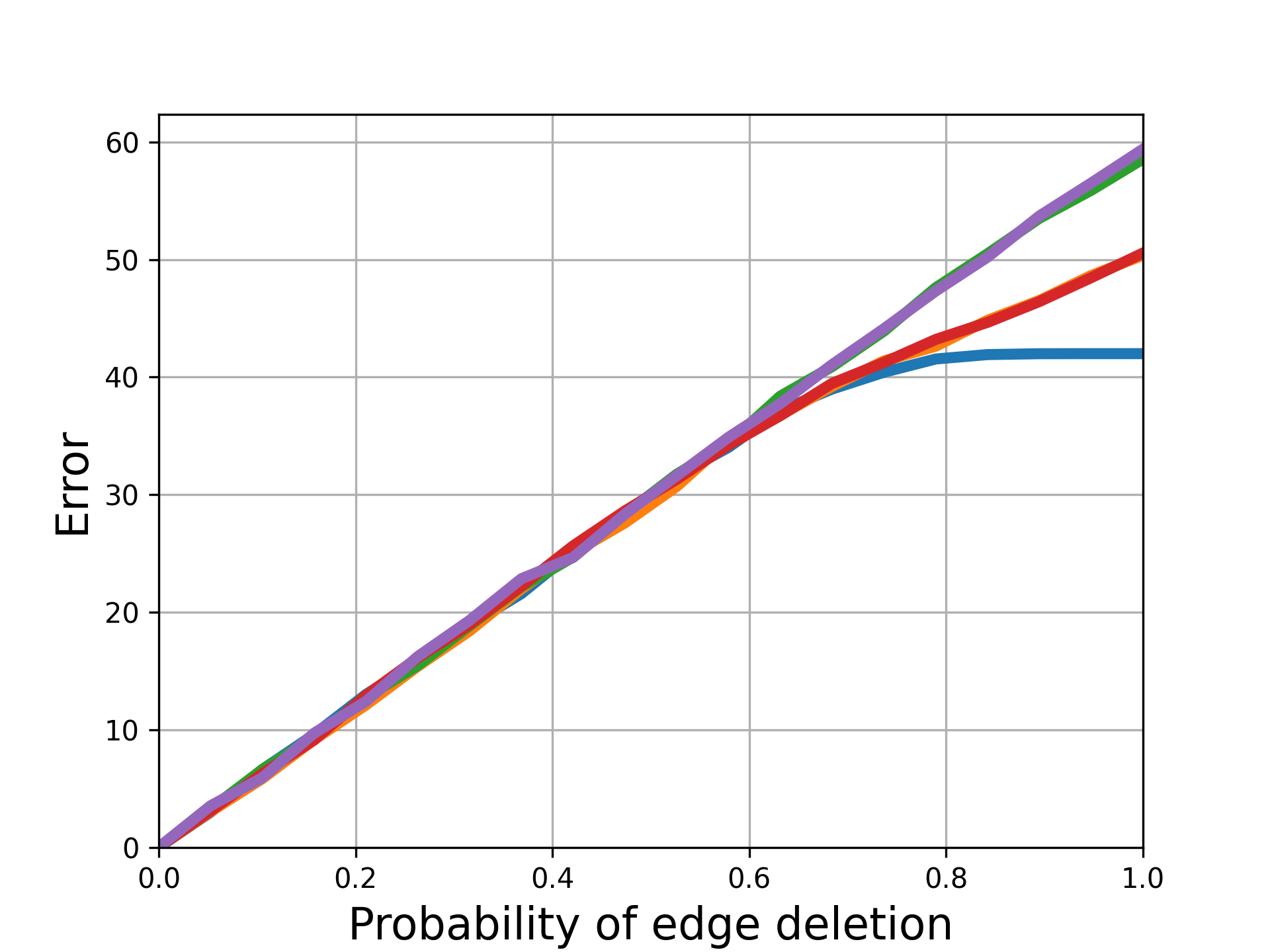}
  \caption{Random edge deletion.}
  \label{fig:exp2_c}
\end{subfigure}
\begin{subfigure}{.24\textwidth}
  \centering
  \includegraphics[width=.999\linewidth]{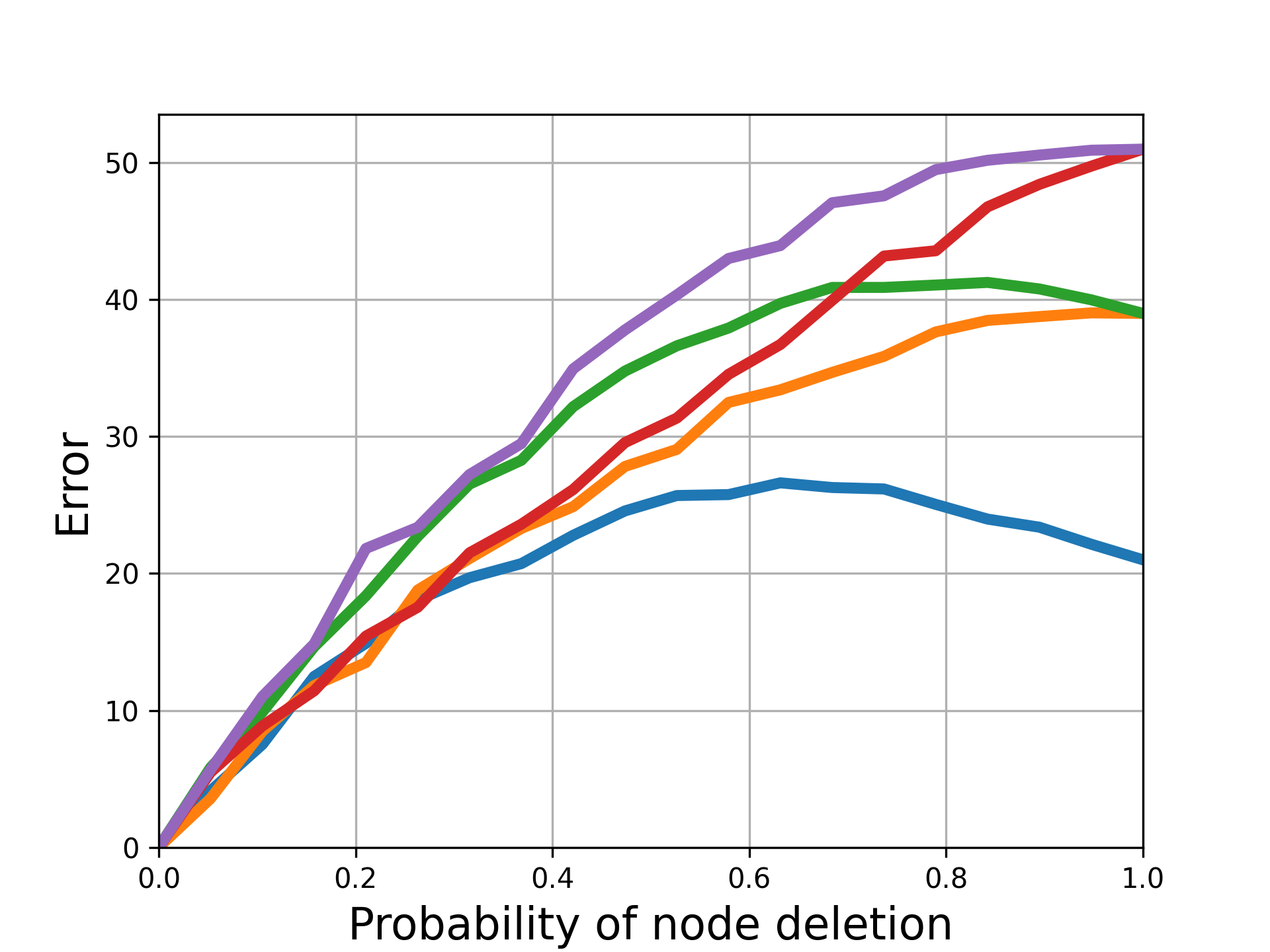}
  \caption{Random node removal.}
  \label{fig:exp2_d}
\end{subfigure}
\caption{Plots of average graph GOSPA metric family with different hyperparameters $\beta$ and $\eta$, comparing the ground truth graph with (a) graphs with random Gaussian noise with increasing noise variance in node attributes; (b) graphs with random edge addition in adjacency matrices; (c) graphs with random edge removal in adjacency matrices; (d) graphs with different sizes.}
\label{fig:exp_compare}
\end{figure}

\subsection{Evaluation of different hyperparameters of the graph GOSPA family}\label{subsec:evaluate hyperparameters}
In this section, we examine the performance of the graph GOSPA metric family with different hyperparameters $\beta$ and $\eta$ to illustrate the higher flexibility of the family compared to the graph GOSPA metric to penalise edge mismatches. We follow the same procedure in Section \ref{subsec:simulations} to create random graphs $X$ and $Y$, but without adding small Gaussian noise to node attributes in cases (b), (c) and (d). In this experiment, we consider $c=3$, $\epsilon=1$, $p=1$ and $\beta \in \{0.3,0.5\}$, $\eta \in \{0.5,0.7\}$. It is worth noticing that the graph GOSPA metric in \cite{gu24} is a special case of the graph GOSPA family, where $\beta=0$ and $\eta=\frac{1}{2}$. 

In Figure \ref{fig:exp2_a}, we see that the curves with the same $\beta$ converge to the same error value as the noise in node attributes increases, but the curve with a larger $\eta$ reaches the saturation value with a lower noise variance level. For curves with the same $\eta$, although they have a different value of saturation, they reach the saturated point at the same noise level in node attributes. 
Figure \ref{fig:exp2_b} shows that the graph GOSPA metrics family with different hyperparameters ($\eta \geq \frac{1}{2}$) produce the same curve when the probability of adding extra edges to graph $Y$ increases. 
The curves in Figure \ref{fig:exp2_c} demonstrate that when more edges have been deleted in graph $Y$, if $\eta > \frac{1}{2}$, the error value will continue to increase, and in this case $\beta$ does not affect the metric value.
Finally, in Figure \ref{fig:exp2_d}, it can be seen that the larger the values of $\eta$ and $\beta$, the higher the error as the cost for the edge mismatches becomes higher.

\subsection{Computation time}\label{subsection: computation time}

In this section, we analyse the computational time of the metrics. We compare both randomly generated graphs $X$ and $Y$ using the same setups in Section \ref{subsec:simulations}, but changing the sizes to $n_X=n_Y\in\{10,20,40,60\}$. Table \ref{tab:time compare undirected} includes the computational times for the graph GOSPA metric family with the LP, MILP and (relaxed) QP implementations. This table also includes the computational times for the original graph GOSPA metric, GED and GCD. 

We can see that GED has the highest computational complexity. Graph GOSPA and GCD are the fastest to compute. Although the graph GOSPA metric family takes longer to compute than the graph GOSPA metric, the LP version of the family is considerably faster to compute than the MILP, which already takes more than 5000 seconds to compute for graphs with 20 nodes. The (relaxed) QP implementation is faster than the MILP implementation, but considerably slower than the LP implementation. Therefore, from a computational standpoint, the LP implementation is the preferred choice for the graph GOSPA metric family.

\begin{table}
    \centering
    \caption{Computational time over undirected graphs (CPU time in seconds).}
    \begin{tabular}{c|c|c|c|c}
    \hline 
    $(n_X,n_Y)$& (10,10) &(20,20) &(40,40) &(60,60) \\
    \hline 
    Graph GOSPA &0.01 &0.11&10.68 &159.72\\
    GED  &28.53 & >5000& - & -\\
    GCD & 0.02& 0.18& 8.75 & 200.17\\
    Graph GOSPA family (LP) & 0.02 &0.30 &28.22& 3282.16 \\
    Graph GOSPA family (MILP) &1.93 & >5000 & - & - \\
    Graph GOSPA family (QP) &14.16 & 304.81&406.09 &>5000 \\
    \hline
    \end{tabular}
    \label{tab:time compare undirected}
\end{table}

\subsection{Experiments on real-world datasets} \label{subsec: molecule experiment}
In this section, we test the proposed graph GOSPA metric family (LP implementation) for classification purposes using it as part of a KNN classifier \cite{cover1967nearest}. In particular, we compare it with KNN classifiers that make use of these metric/distances: the graph GOSPA metric, the GCD and the GED. 

We consider these real-world data sets: MUTAG \cite{Morris+2020}, the Letter Database \cite{riesen2008iam}, and the PTC (FM), PTC (MM) and PTC (MR) datasets \cite{helma2001predictive}. The MUTAG dataset is a dataset that contains 188 molecules from two classes based on the mutagenic effect on a bacterium. Each molecule in the dataset contains around 18 nodes and 20 edges. The Letter Database contains 2250 graphs for 15 capital letters of the Roman alphabet that consist of straight lines only, and three levels of distortions are applied to the dataset (low, medium and high). As the accuracy in the dataset with small distortions is almost 100\%, we use the dataset with medium (letter-m) and high distortions (letter-h) to test the proposed method. The predictive toxicology challenge (PTC) dataset contains 346 compounds labelled according to their carcinogenicity in rats and mice. The PTC dataset contains several subsets, and we use the data based on female mice, male mice and male rats in this paper.

The training and testing data are set to an 80/20 ratio. For the graph GOSPA metric family, we set $p=1$. The rest of the hyperparameters are determined by first selecting a grid on uniform points in these intervals, $c\in[1,10]$, $\epsilon\in [0.1,5]$, $\beta\in[0,1]$, $\eta\in[0,1]$, and then using random search cross-validation \cite{Bergstra12}. For the graph GOSPA metric, we follow the same approach and intervals to sample $c$ and $\epsilon$. For GCD, the considered intervals for the parameters are $c\in[1,10]$ and $\lambda\in[0.1,5]$. For GED, the interval for $c$ is $[1,10]$. Additionally, in some cases, the GED takes too long to compute, so we set the time limit for each GED computation to 10 seconds.
The optimal parameter combinations are selected based on 5-fold cross-validation on training sets for each baseline, and then tested on testing sets. 

The classification accuracy of these methods is provided in Table \ref{tab:molecule dataset}.
\begin{table*}[t]
    \centering
    \caption{Classification accuracy of the KNN classifiers on real-world datasets.}
    \begin{tabular}{c|c|c|c|c|c|c}
    \hline 
        &MUTAG & Letter-m &Letter-h&PTC (FM)& PTC (MM)& PTC (MR) \\
    \hline
    Graph GOSPA & \textbf{89.47\%} &94.33\% &\textbf{91.78\%}& 59.47\% & 57.35\%& 47.83\%\\
    GCD  & 86.84\% &93.94\% & 88.44\%& 57.79\% &52.94\% & 56.52\%\\
    GED  & 84.21\% & 89.33\%& 84.22\%& 58.77\% & 58.82\% & 52.17\%\\ 
    Graph GOSPA family & \textbf{89.47\%} & \textbf{95.78\%}& \textbf{91.78\%} &\textbf{64.29\%} & \textbf{60.29\%}& \textbf{57.94\%} \\
    \hline
    \end{tabular}
    \label{tab:molecule dataset}
\end{table*}
These results demonstrate that the graph GOSPA metric family performs the best in all datasets, though graph GOSPA is equally good in the MUTAG and Letter-high datasets. Since the graph GOSPA metric is a special case of the graph GOSPA family with $\beta=0$ and $\eta=0.5$, they can sometimes have similar results. In the Letter-medium, PTC (FM), PTC (MM) and PTC (MR) datasets, the graph GOSPA metric family has better results than other methods, which indicates that for these tasks, the extra hyperparameters $\beta$ and $\eta$ in the graph GOSPA metric family can offer more flexibility to fit the data better.

In Table \ref{tab:opt_hyperparam}, we show the optimal hyperparameters obtained for the graph GOSPA metric and the graph GOSPA family. For Letter-m and Letter-h datasets, the parameters $c$ and $\epsilon$ for both metrics are similar, and they have similar performance of these datasets. For MUTAG, PTC (FM), PTC (MM) and PTC (MR) datasets, the optimal hyperparameters of the graph GOSPA family are considerably different from the hyperparameters in the graph GOSPA metric. In MUTAG, their performance is similar, but for the other datasets, the graph GOSPA family performs better.

\begin{table}
    \centering
    \caption{Optimal hyperparameter obtained.}
    \begin{tabular}{c|c|c}
    \hline 
        &Graph GOSPA& Graph GOSPA family \\
        &($c$, $\epsilon$, $\eta=0.5$, $\beta=0$) &($c$, $\epsilon$, $\eta$, $\beta$)  \\
    \hline
    MUTAG& (6.0, 1.73)& (3.0, 1.2, 0.8, 0.78)\\
    Letter-m& (2.0, 1.19) & (2.0, 1.2, 0.55, 0.67)\\
    Letter-h& (3.0, 1.19)& (3.0, 1.2, 0.8, 0.73)\\
    PTC (FM)& (9.0, 0.1)& (6.0, 2.3, 0.03, 0.17)\\
    PTC (MM)& (9.0, 1.73) & (1.0, 1.2, 0.86, 0.5)\\
    PTC (MR)& (9.0, 0.1) & (10.0, 0.64, 0.78, 0.25)\\
    \hline
    \end{tabular}
    \label{tab:opt_hyperparam}
\end{table}

\section{Conclusion}\label{sec:conclusion}

In this paper, we have proposed a family of graph GOSPA metrics which are capable of comparing undirected graphs with different sizes in a clearly interpretable manner. The graph GOSPA metric family with binary assignments meets the metric properties for undirected graphs with node attributes and different sizes. 
We show that the graph GOSPA metric family can be computed by solving a quadratic programming problem. With the linearisation of the quadratic programming problem, we can use linear programming to find an approximate solution. 

The graph GOSPA metric family increases the flexibility of the graph GOSPA metric to penalise edge mismatch penalties, enabling the penalty of unassigned edges and a range of penalties for half-assigned edge mismatches. It nevertheless requires increased computation compared to the graph GOSPA metric. A line of future work is the development of computational tools to compute the metric for large scale graphs, for instance, using the alternating direction method of multipliers \cite{Moharrer20}.

\bibliographystyle{IEEEtran}
\bibliography{references}

\clearpage

\title{}

\maketitle
{\LARGE Supplementary material: A family of graph GOSPA metrics for graphs with different sizes}{\LARGE\par}

\appendices

\setcounter{equation}{0}
\renewcommand{\theequation}{S.\arabic{equation}}

\section{}\label{app:proof of triangle inequality}

In this appendix, we give the proof of the triangle inequality of the graph GOSPA metric family in \eqref{eq:gospa family defn}. We can write \eqref{eq:gospa family defn} as a function of the assignment set
as $d\left(X,Y,\gamma\right)$, where we have omitted the superscripts for notational clarity. We assume
we have three graphs $X$, $Y$ and $Z$. Let $\gamma_{XY}^{*}$,
$\gamma_{XZ}^{*}$ and $\gamma_{ZY}^{*}$ be the assignment sets that
minimise $d\left(X,Y,\gamma\right)$,
$d\left(X,Z,\gamma\right)$ and $d\left(Z,Y,\gamma\right)$.
Therefore, the triangle inequality property is met if the following
inequality is met for any $X$, $Y$ and $Z$
\begin{align}
d\left(X,Y,\gamma_{XY}^{*}\right) & \leq d\left(X,Z,\gamma_{XZ}^{*}\right)+d\left(Z,Y,\gamma_{ZY}^{*}\right).
\end{align}

Given $\gamma_{XZ}^{*}$ and $\gamma_{ZY}^{*}, $we construct an assignment
set $\gamma_{XY}$ between $X$ and $Y$ such that the nodes in $X$
and $Y$ are assigned if and only if they are assigned through the
assignment sets in $Z$. That is
\begin{align}
\gamma_{XY} & =\left\{ \left(i,j\right):\exists k:\left(i,k\right)\in\gamma_{XZ}^{*}\,\mathrm{and}\,\left(k,j\right)\in\gamma_{ZY}^{*}\right\}. \label{eq:assignment_XY}
\end{align}

The triangle inequality is met if we prove 
\begin{align}
d\left(X,Y,\gamma_{XY}\right) & \leq d\left(X,Z,\gamma_{XZ}^{*}\right)+d\left(Z,Y,\gamma_{ZY}^{*}\right).
\end{align}
To prove the previous equation, we show that for any $\gamma_{XY}$,
$\gamma_{XZ}$ and $\gamma_{ZY}$ that meet (\ref{eq:assignment_XY}), the following result holds
\begin{align}
d\left(X,Y,\gamma_{XY}\right) & \leq d\left(X,Z,\gamma_{XZ}\right)+d\left(Z,Y,\gamma_{ZY}\right).\label{eq:triangle_ineq_fixed_assignment}
\end{align}

There are two aspects that make this proof more challenging than the one in \cite{gu24}: it makes direct use of assignment sets instead of  matrices and it must determine the suitable range of values of $\beta$ and $\eta$. We first define the following notation for the proof. Given $\gamma_{XY}$,
$\gamma_{XZ}$ and $\gamma_{ZY}$ that meet (\ref{eq:assignment_XY}),
we define the subsets $\gamma_{XZ}^{1}\subseteq\gamma_{XZ}$ and $\gamma_{ZY}^{1}\subseteq\gamma_{ZY}$
that include assignments that connect nodes from $X$ to $Y$ through
$Z$. That is, we can write 
\begin{equation}
\gamma_{XY}=\left\{ \left(i,j\right):\exists k:\left(i,k\right)\in\gamma_{XZ}^{1}\,\mathrm{and}\,\left(k,j\right)\in\gamma_{ZY}^{1}\right\}. 
\end{equation}
This implies that $|\gamma_{XY}|=|\gamma_{XZ}^{1}|=|\gamma_{ZY}^{1}|$. We also define $\gamma_{XZ}^{0}$ and $\gamma_{ZY}^{0}$ as
the sets of assignments that are not part of an assignment connecting
$X$ and $Y$ through $Z$ implying
\begin{flalign}
\gamma_{XZ} & =\gamma_{XZ}^{0}\cup\gamma_{XZ}^{1}\\
\gamma_{ZY} & =\gamma_{ZY}^{0}\cup\gamma_{ZY}^{1}.
\end{flalign}
Finally, we use the notation $\gamma_{XY}=\gamma_{XZ}^{1}\circ\gamma_{ZY}^{1}$
to represent that the assignments from $X$ to $Y$ are composed by
the assignments from $X$ to $Z$ in $\gamma_{XZ}^{1}$ and the assignments
from $Z$ to $Y$ in $\gamma_{ZY}^{1}$.

The proof of \eqref{eq:triangle_ineq_fixed_assignment} can be divided into two separate parts:
1) The first one is the proof of the triangle inequality for the localisation cost (including missed and false nodes). That is, we want 
\begin{flalign}
 & \sum_{\left(i,j\right)\in\gamma_{XY}}d^{p}\left(x_{i},y_{j}\right)+\frac{c^{p}}{2}\left(\left|V_X\right|+\left|V_Y\right|-2\left|\gamma_{XY}\right|\right)\nonumber \\
 & \leq\sum_{\left(i,k\right)\in\gamma_{XZ}^{1},\left(k,j\right)\in\gamma_{ZY}^{1}}\left(d\left(x_{i},z_{k}\right)+d\left(z_{k},y_{j}\right)\right)^{p}\nonumber \\
 & +\sum_{\left(i,k\right)\in\gamma_{XZ}^{0}}d^{p}\left(x_{i},z_{k}\right)+\sum_{\left(k,j\right)\in\gamma_{ZY}^{0}}d^{p}\left(z_{k},y_{j}\right)\nonumber \\
 & +\frac{c^{p}}{2}\left(\left|V_X\right|+|V_Z|-2\left|\gamma_{XZ}\right|\right)\nonumber \\
 & +\frac{c^{p}}{2}\left(|V_Z|+\left|V_Y\right|-2\left|\gamma_{ZY}\right|\right).\label{eq:loc_cost_inequality}
\end{flalign}
This part of the proof is provided in Appendix \ref{proof:loc cost triangle}. 

2) The second part of the proof is the proof of the triangle inequality for the edge cost (to the $p$-th power):
\begin{flalign}
    e_{XY}^{p}\left(\gamma_{XY}\right) & \leq e_{XZ}^{p}\left(\gamma_{XZ}\right)+e_{ZY}^{p}\left(\gamma_{ZY}\right).\label{eq:edge_cost_inequality}
\end{flalign}
This result is proved in Appendix \ref{proof:edge cost triangle}.

To prove \eqref{eq:triangle_ineq_fixed_assignment}, we start with $d\left(X,Y,\gamma_{XY}\right)$ and apply the inequalities in \eqref{eq:loc_cost_inequality} and \eqref{eq:edge_cost_inequality}
\begin{flalign}
&d\left(X,Y,\gamma_{XY}\right) =\left(\sum_{\left(i,j\right)\in\gamma_{XY}}d^{p}\left(x_{i},y_{j}\right) \right. \nonumber \\
& \left.+\frac{c^{p}}{2}\left(\left|V_X\right|+\left|V_Y\right|-2\left|\gamma_{XY}\right|\right)+e_{XY}^{p}\left(\gamma_{XY}\right)\right)^{1/p} \nonumber \\
 & \leq\Biggl(\sum_{\left(i,k\right)\in\gamma_{XZ}^{1},\left(k,j\right)\in\gamma_{ZY}^{1}}\left(d\left(x_{i},z_{k}\right)+d\left(z_{k},y_{j}\right)\right)^{p} \nonumber \\
 &+\sum_{\left(i,k\right)\in\gamma_{XZ}^{0}}d^{p}\left(x_{i},z_{k}\right)+\sum_{\left(k,j\right)\in\gamma_{ZY}^{0}}d^{p}\left(z_{k},y_{j}\right)  \nonumber \\
 &  +\frac{c^{p}}{2}\left(\left|V_X\right|+|V_Z|-2\left|\gamma_{XZ}\right|\right) \nonumber \\
 & +\frac{c^{p}}{2}\left(|V_Z|+\left|V_Y\right|-2\left|\gamma_{ZY}\right|\right)\nonumber\\
 &  +e_{XZ}^{p}\left(\gamma_{XZ}\right)+e_{ZY}^{p}\left(\gamma_{ZY}\right)\Biggl)^{1/p}. \label{eq:full_triangle_inequality}
\end{flalign}
Then, we apply Minkowski's inequality see \cite[Eq.(3)]{Rahmathullah17} and \cite{kubrusly2011elements} to obtain
\begin{flalign}
&d\left(X,Y,\gamma_{XY}\right) \nonumber \\
& \leq\left(\sum_{\left(i,k\right)\in\gamma_{XZ}^{1},\left(k,j\right)\in\gamma_{ZY}^{1}}d^{p}\left(x_{i},z_{k}\right)+\sum_{\left(i,k\right)\in\gamma_{XZ}^{0}}d^{p}\left(x_{i},z_{k}\right)\right. \nonumber \\
 & \left.+\frac{c^{p}}{2}\left(\left|V_X\right|+|V_Z|-2\left|\gamma_{XZ}\right|\right)+e_{XZ}^{p}\left(\gamma_{XZ}\right)\right)^{1/p} \nonumber\\
 & +\left(\sum_{\left(i,k\right)\in\gamma_{XZ}^{1},\left(k,j\right)\in\gamma_{ZY}^{1}}d^{p}\left(z_{k},y_{j}\right)+\sum_{\left(k,j\right)\in\gamma_{ZY}^{0}}d^{p}\left(z_{k},y_{j}\right)\right. \nonumber\\
 & \left.+\frac{c^{p}}{2}\left(|V_Z|+\left|V_Y\right|-2\left|\gamma_{ZY}\right|\right)+e_{ZY}^{p}\left(\gamma_{ZY}\right)\right)^{1/p} \nonumber \\
 & =d\left(X,Z,\gamma_{XZ}\right)+d\left(Z,Y,\gamma_{ZY}\right). \label{eq:triangle_inequality_final_form}
\end{flalign}

This finishes the proof of \eqref{eq:triangle_ineq_fixed_assignment}.

\subsection{Localisation cost inequality}\label{proof:loc cost triangle}
In this section,  we prove \eqref{eq:loc_cost_inequality}. On the left-hand side of \eqref{eq:loc_cost_inequality}, we have 

\begin{flalign}
&\sum_{\left(i,j\right)\in\gamma_{XY}}d^{p}\left(x_{i},y_{j}\right)+\frac{c^{p}}{2}\left(\left|V_X\right|+\left|V_Y\right|-2\left|\gamma_{XY}\right|\right) \nonumber \\
  =&\sum_{\left(i,j\right)\in\gamma_{XZ}^{1}\circ\gamma_{ZY}^{1}}d^{p}\left(x_{i},y_{j}\right)+\frac{c^{p}}{2}\left(\left|V_X\right|+\left|V_Y\right|-2\left|\gamma_{XY}\right|\right)\nonumber\\
  \leq&\sum_{\left(i,k\right)\in\gamma_{XZ}^{1},\left(k,j\right)\in\gamma_{ZY}^{1}}\left(d\left(x_{i},z_{k}\right)+d\left(z_{k},y_{j}\right)\right)^{p}\nonumber\\
 &+\frac{c^{p}}{2}\left(\left|V_X\right|+\left|V_Y\right|-2\left|\gamma_{XZ}^{1}\circ\gamma_{ZY}^{1}\right|\right)\nonumber\\ \leq&\sum_{\left(i,k\right)\in\gamma_{XZ}^{1},\left(k,j\right)\in\gamma_{ZY}^{1}}\left(d\left(x_{i},z_{k}\right)+d\left(z_{k},y_{j}\right)\right)^{p}\nonumber\\
 &+\frac{c^{p}}{2}\left(\left|V_X\right|+\left|V_Y\right|-2\left|\gamma_{XZ}\right|-2\left|\gamma_{ZY}\right|\right)\nonumber\\
 \leq&\sum_{\left(i,k\right)\in\gamma_{XZ}^{1},\left(k,j\right)\in\gamma_{ZY}^{1}}\left(d\left(x_{i},z_{k}\right)+d\left(z_{k},y_{j}\right)\right)^{p}\nonumber\\
 & +\frac{c^{p}}{2}\left(\left|V_X\right|+|V_Z|-2\left|\gamma_{XZ}\right|\right)+\frac{c^{p}}{2}\left(|V_Z|+\left|V_Y\right|-2\left|\gamma_{ZY}\right|\right)\nonumber\\
 \leq&\sum_{\left(i,k\right)\in\gamma_{XZ}^{1},\left(k,j\right)\in\gamma_{ZY}^{1}}\left(d\left(x_{i},z_{k}\right)+d\left(z_{k},y_{j}\right)\right)^{p}\nonumber\\
 & +\sum_{\left(i,k\right)\in\gamma_{XZ}^{0}}d^{p}\left(x_{i},z_{k}\right)+\sum_{\left(k,j\right)\in\gamma_{ZY}^{0}}d^{p}\left(z_{k},y_{j}\right)\nonumber\\
 & +\frac{c^{p}}{2}\left(\left|V_X\right|+|V_Z|-2\left|\gamma_{XZ}\right|\right) +\frac{c^{p}}{2}\left(|V_Z|+\left|V_Y\right|-2\left|\gamma_{ZY}\right|\right). \label{eq:proof_local_cost_1} 
\end{flalign}
Thus, the inequality in \eqref{eq:loc_cost_inequality} has been proved.

\subsection{Inequality for edge cost}\label{proof:edge cost triangle}
In this section, we prove \eqref{eq:edge_cost_inequality}. The edge mismatch cost for $X$ and $Y$ can be written as
\begin{flalign}
    &e_{XY}^{p}\left(\gamma_{XY}\right) \nonumber \\
    =& \sum_{\left(i_{1},j_{1}\right)\in\gamma_{XZ}^{1}\circ\gamma_{ZY}^{1}}\sum_{\left(i_{2},j_{2}\right)\in\gamma_{XZ}^{1}\circ\gamma_{ZY}^{1}}\frac{\epsilon^{p}}{2}|A_{X}(i_{1},i_{2})-A_{Y}(j_{1},j_{2})| \nonumber \\
 &+ \eta\epsilon^{p}\sum_{\left(i_{1},j_{1}\right)\in\gamma_{XZ}^{1}\circ\gamma_{ZY}^{1}}\nonumber \\
 &\left[\sum_{i_{2}\in U_{X}\left(\gamma_{XZ}\right)}A_{X}(i_{1},i_{2}) + \sum_{\left(i_{2},k_{2}\right)\in\gamma_{XZ}^{0}}A_{X}(i_{1},i_{2})\right] \nonumber \\
  &+\eta\epsilon^{p}\sum_{\left(i_{1},j_{1}\right)\in\gamma_{XZ}^{1}\circ\gamma_{ZY}^{1}} \nonumber \\
  &\left[\sum_{j_{2}\in U_{Y}\left(\gamma_{ZY}\right)}A_{Y}(j_{1},j_{2}) + \sum_{\left(k_{2},j_{2}\right)\in\gamma_{ZY}^{0}}A_{Y}(j_{1},j_{2})\right] \nonumber\\
 &+\frac{\beta}{2}\epsilon^{p}\sum_{i_{1}\in U_{X}\left(\gamma_{XZ}\right)}\sum_{i_{2}\in U_{X}\left(\gamma_{XZ}\right)}A_{X}(i_{1},i_{2}) \nonumber \\
 &+\beta\epsilon^{p}\sum_{i_{1}\in U_{X}\left(\gamma_{XZ}\right)}\sum_{\left(i_{2},k_{2}\right)\in\gamma_{XZ}^{0}}A_{X}(i_{1},i_{2})\nonumber \\
 &+ \frac{\beta}{2}\epsilon^{p}\sum_{\left(i_{1},k_{1}\right)\in\gamma_{XZ}^{0}}\sum_{\left(i_{2},k_{2}\right)\in\gamma_{XZ}^{0}}A_{X}(i_{1},i_{2}) \nonumber\\
 &+ \frac{\beta}{2}\epsilon^{p}\sum_{j_{1}\in U_{Y}\left(\gamma_{ZY}\right)}\sum_{j_{2}\in U_{Y}\left(\gamma_{ZY}\right)}A_{Y}(j_{1},j_{2}) \nonumber \\
 &+ \beta\epsilon^{p}\sum_{j_{1}\in U_{Y}\left(\gamma_{ZY}\right)}\sum_{\left(k_{2},j_{2}\right)\in\gamma_{ZY}^{0}}A_{Y}(j_{1},j_{2})\nonumber\\
 &+\frac{\beta}{2}\epsilon^{p}\sum_{\left(k_{1},j_{1}\right)\in\gamma_{ZY}^{0}}\sum_{\left(k_{2},j_{2}\right)\in\gamma_{ZY}^{0}}A_{Y}(j_{1},j_{2}).\label{eq:edge_cost_inequality_proof}
\end{flalign}
Now, we operate based on the types of costs in \eqref{eq:edge_cost_inequality_proof}.
\subsubsection{Cost for assigned edges}
The cost for the assigned edges in \eqref{eq:edge_cost_inequality_proof} meets
\begin{flalign}
     & \frac{\epsilon^{p}}{2}\sum_{\left(i_{1},j_{1}\right)\in\gamma_{XZ}^{1}\circ\gamma_{ZY}^{1}}\sum_{\left(i_{2},j_{2}\right)\in\gamma_{XZ}^{1}\circ\gamma_{ZY}^{1}} \nonumber \\
    & |A_{X}(i_{1},i_{2})-A_{Y}(j_{1},j_{2})| \nonumber \\
  =&\frac{\epsilon^{p}}{2}\sum_{\left(i_{1},k_{1}\right)\in\gamma_{XZ}^{1},\left(k_{1},j_{1}\right)\in\gamma_{ZY}^{1}}\sum_{\left(i_{2},k_{2}\right)\in\gamma_{XZ}^{1},(k_{2},j_{2})\in\gamma_{ZY}^{1}} \nonumber \\
  &|A_{X}(i_{1},i_{2})-A_{Y}(j_{1},j_{2})+A_{Z}(k_{1},k_{2})-A_{Z}(k_{1},k_{2})|\nonumber\\
 \leq&\frac{\epsilon^{p}}{2}\sum_{\left(i_{1},k_{1}\right)\in\gamma_{XZ}^{1},\left(k_{1},j_{1}\right)\in\gamma_{ZY}^{1}}\sum_{\left(i_{2},k_{2}\right)\in\gamma_{XZ}^{1},(k_{2},j_{2})\in\gamma_{ZY}^{1}} \nonumber \\ 
 &\left[|A_{X}(i_{1},i_{2})-A_{Z}(k_{1},k_{2})|+|A_{Z}(k_{1},k_{2})-A_{Y}(j_{1},j_{2})|\right]\nonumber\\
 \leq&\frac{\epsilon^{p}}{2}\sum_{\left(i_{1},k_{1}\right)\in\gamma_{XZ}^{1},}\sum_{\left(i_{2},k_{2}\right)\in\gamma_{XZ}^{1}}|A_{X}(i_{1},i_{2})-A_{Z}(k_{1},k_{2})| \nonumber \\
 &+\frac{\epsilon^{p}}{2}\sum_{\left(k_{1},j_{1}\right)\in\gamma_{ZY}^{1}}\sum_{(k_{2},j_{2})\in\gamma_{ZY}^{1}}|A_{Z}(k_{1},k_{2})-A_{Y}(j_{1},j_{2})|. \label{eq:assigned_edge_cost}
\end{flalign}
\subsubsection{Cost for half-assigned edges}
We first consider the terms for the half-assigned edges in $X$ in \eqref{eq:edge_cost_inequality_proof} to yield the inequality.
\begin{flalign}
     & \eta\epsilon^{p}\sum_{\left(i_{1},j_{1}\right)\in\gamma_{XZ}^{1}\circ\gamma_{ZY}^{1}} \nonumber \\
     &\left[\sum_{i_{2}\in U_{X}\left(\gamma_{XZ}\right)}A_{X}(i_{1},i_{2})+\sum_{\left(i_{2},k_{2}\right)\in\gamma_{XZ}^{0}}A_{X}(i_{1},i_{2})\right] \nonumber \\
 =& \eta\epsilon^{p}\sum_{\left(i_{1},j_{1}\right)\in\gamma_{XZ}^{1}\circ\gamma_{ZY}^{1}}\sum_{i_{2}\in U_{X}\left(\gamma_{XZ}\right)}A_{X}(i_{1},i_{2})\nonumber \\
 &+ \eta\epsilon^{p}\sum_{\left(i_{1},k_{1}\right)\in\gamma_{XZ}^{1},(k_{1},j_{1})\in\gamma_{ZY}^{1}}\sum_{\left(i_{2},k_{2}\right)\in\gamma_{XZ}^{0}}A_{X}(i_{1},i_{2}) \nonumber \\
  =&\eta\epsilon^{p}\sum_{\left(i_{1},k_{1}\right)\in\gamma_{XZ}^{1}}\sum_{i_{2}\in U_{X}\left(\gamma_{XZ}\right)}A_{X}(i_{1},i_{2}) \nonumber \\
 & +\eta\epsilon^{p}\sum_{\left(i_{1},k_{1}\right)\in\gamma_{XZ}^{1}}\sum_{\left(i_{2},k_{2}\right)\in\gamma_{XZ}^{0}}A_{X}(i_{1},i_{2})\nonumber \\
  =&\eta\epsilon^{p}\sum_{\left(i_{1},k_{1}\right)\in\gamma_{XZ}^{1}}\sum_{i_{2}\in U_{X}\left(\gamma_{XZ}\right)}A_{X}(i_{1},i_{2})\nonumber\\
 & +\eta\epsilon^{p}\sum_{\left(i_{1},k_{1}\right)\in\gamma_{XZ}^{1}}\sum_{\left(i_{2},k_{2}\right)\in\gamma_{XZ}^{0}} \nonumber\\
 &|A_{X}(i_{1},i_{2})-A_{Z}(k_{1},k_{2})+A_{Z}(k_{1},k_{2})|\nonumber\\
 \leq&\eta\epsilon^{p}\sum_{\left(i_{1},k_{1}\right)\in\gamma_{XZ}^{1}}\sum_{i_{2}\in U_{X}\left(\gamma_{XZ}\right)}A_{X}(i_{1},i_{2})\nonumber\\
 & +\eta\epsilon^{p}\sum_{\left(i_{1},k_{1}\right)\in\gamma_{XZ}^{1}}\sum_{\left(i_{2},k_{2}\right)\in\gamma_{XZ}^{0}}|A_{X}(i_{1},i_{2})-A_{Z}(k_{1},k_{2})|\nonumber\\
 & +\eta\epsilon^{p}\sum_{\left(i_{1},k_{1}\right)\in\gamma_{XZ}^{1}}\sum_{\left(i_{2},k_{2}\right)\in\gamma_{XZ}^{0}}A_{Z}(k_{1},k_{2})\nonumber\\
 \leq&\eta\epsilon^{p}\sum_{\left(i_{1},k_{1}\right)\in\gamma_{XZ}^{1}}\sum_{i_{2}\in U_{X}\left(\gamma_{XZ}\right)}A_{X}(i_{1},i_{2})\nonumber\\
 & +\eta\epsilon^{p}\sum_{\left(i_{1},k_{1}\right)\in\gamma_{XZ}^{1}}\sum_{\left(i_{2},k_{2}\right)\in\gamma_{XZ}^{0}}|A_{X}(i_{1},i_{2})-A_{Z}(k_{1},k_{2})|\nonumber\\
 & +\eta\epsilon^{p}\sum_{\left(k_{1},j_{1}\right)\in\gamma_{ZY}^{1}}\sum_{k_{2}\in U_{Z}\left(\gamma_{ZY}\right)}A_{Z}(k_{1},k_{2}). \label{eq:half-assigned_edge_cost_X}
\end{flalign}
Analogously, for the half-assigned edges in $Y$, we obtain
\begin{flalign}
     & \eta\epsilon^{p}\sum_{\left(i_{1},j_{1}\right)\in\gamma_{XZ}^{1}\circ\gamma_{ZY}^{1}} \nonumber \\
     &\left[\sum_{j_{2}\in U_{Y}\left(\gamma_{ZY}\right)}A_{Y}(j_{1},j_{2})+\sum_{\left(k_{2},j_{2}\right)\in\gamma_{ZY}^{0}}A_{Y}(j_{1},j_{2})\right] \nonumber \\
 \leq&\eta\epsilon^{p}\sum_{\left(k_{1},j_{1}\right)\in\gamma_{ZY}^{1}}\sum_{j_{2}\in U_{Y}\left(\gamma_{ZY}\right)}A_{Y}(j_{1},j_{2}) \nonumber \\
 & +\eta\epsilon^{p}\sum_{\left(k_{1},j_{1}\right)\in\gamma_{ZY}^{1}}\sum_{\left(k_{2},j_{2}\right)\in\gamma_{ZY}^{0}}\text{|}A_{Y}(j_{1},j_{2})-A_{Z}(k_{1},k_{2})| \nonumber \\
 & +\eta\epsilon^{p}\sum_{\left(i_{1},k_{1}\right)\in\gamma_{XZ}^{1}}\sum_{k_{2}\in U_{Z}\left(\gamma_{XZ}\right)}A_{Z}(k_{1},k_{2}). \label{eq:half-assigned_edge_cost_Y}
\end{flalign}

\subsubsection{Cost for unassigned edges}
Operating with the cost for unassigned edges in $X$, we obtain the following inequality
\begin{flalign}
     & \frac{\beta}{2}\epsilon^{p}\sum_{i_{1}\in U_{X}\left(\gamma_{XZ}\right)}\sum_{i_{2}\in U_{X}\left(\gamma_{XZ}\right)}A_{X}(i_{1},i_{2})\nonumber\\
 & +\beta\epsilon^{p}\sum_{i_{1}\in U_{X}\left(\gamma_{XZ}\right)}\sum_{\left(i_{2},k_{2}\right)\in\gamma_{XZ}^{0}}A_{X}(i_{1},i_{2})\nonumber\\
 & \frac{\beta}{2}\beta\epsilon^{p}\sum_{\left(i_{1},k_{1}\right)\in\gamma_{XZ}^{0}}\sum_{\left(i_{2},k_{2}\right)\in\gamma_{XZ}^{0}}A_{X}(i_{1},i_{2})\nonumber\\
 =&\frac{\beta}{2}\epsilon^{p}\sum_{i_{1}\in U_{X}\left(\gamma_{XZ}\right)}\sum_{i_{2}\in U_{X}\left(\gamma_{XZ}\right)}A_{X}(i_{1},i_{2})\nonumber\\
 & +\beta\epsilon^{p}\sum_{i_{1}\in U_{X}\left(\gamma_{XZ}\right)}\sum_{\left(i_{2},k_{2}\right)\in\gamma_{XZ}^{0}}A_{X}(i_{1},i_{2})\nonumber\\
 & +\frac{\beta}{2}\epsilon^{p}\sum_{\left(i_{1},k_{1}\right)\in\gamma_{XZ}^{0}}\sum_{\left(i_{2},k_{2}\right)\in\gamma_{XZ}^{0}}\nonumber\\
 &|A_{X}(i_{1},i_{2})-A_{Z}(k_{1},k_{2})+A_{Z}(k_{1},k_{2})|\nonumber\\
  \leq&\frac{\beta}{2}\epsilon^{p}\sum_{i_{1}\in U_{X}\left(\gamma_{XZ}\right)}\sum_{i_{2}\in U_{X}\left(\gamma_{XZ}\right)}A_{X}(i_{1},i_{2})\nonumber\\
 & +\beta\epsilon^{p}\sum_{i_{1}\in U_{X}\left(\gamma_{XZ}\right)}\sum_{\left(i_{2},k_{2}\right)\in\gamma_{XZ}^{0}}A_{X}(i_{1},i_{2})\nonumber\\
 & +\frac{\beta}{2}\epsilon^{p}\sum_{\left(i_{1},k_{1}\right)\in\gamma_{XZ}^{0}}\sum_{\left(i_{2},k_{2}\right)\in\gamma_{XZ}^{0}}|A_{X}(i_{1},i_{2})-A_{Z}(k_{1},k_{2})|\nonumber\\
 & +\frac{\beta}{2}\epsilon^{p}\sum_{\left(i_{1},k_{1}\right)\in\gamma_{XZ}^{0}}\sum_{\left(i_{2},k_{2}\right)\in\gamma_{XZ}^{0}}A_{Z}(k_{1},k_{2})\nonumber\\
 \leq&\frac{\beta}{2}\epsilon^{p}\sum_{i_{1}\in U_{X}\left(\gamma_{XZ}\right)}\sum_{i_{2}\in U_{X}\left(\gamma_{XZ}\right)}A_{X}(i_{1},i_{2})\nonumber\\
 & +\beta\epsilon^{p}\sum_{i_{1}\in U_{X}\left(\gamma_{XZ}\right)}\sum_{\left(i_{2},k_{2}\right)\in\gamma_{XZ}^{0}}A_{X}(i_{1},i_{2})\nonumber\\
 & +\frac{\beta}{2}\epsilon^{p}\sum_{\left(i_{1},k_{1}\right)\in\gamma_{XZ}^{0}}\sum_{\left(i_{2},k_{2}\right)\in\gamma_{XZ}^{0}}|A_{X}(i_{1},i_{2})-A_{Z}(k_{1},k_{2})|\nonumber\\
 & +\frac{\beta}{2}\epsilon^{p}\sum_{k_{1}\in U_{Z}\left(\gamma_{ZY}\right)}\sum_{k_{2}\in U_{Z}\left(\gamma_{ZY}\right)}A_{Z}(k_{1},k_{2}). \label{eq:unassigned_edge_cost_X}
\end{flalign}
Then we proceed analogously for the unassigned edges in $Y$ to obtain the inequality
\begin{flalign}
     & \frac{\beta}{2}\epsilon^{p}\sum_{j_{1}\in U_{Y}\left(\gamma_{ZY}\right)}\sum_{j_{2}\in U_{Y}\left(\gamma_{ZY}\right)}A_{Y}(j_{1},j_{2})\nonumber\\
 & +\beta\epsilon^{p}\sum_{j_{1}\in U_{Y}\left(\gamma_{ZY}\right)}\sum_{\left(k_{2},j_{2}\right)\in\gamma_{ZY}^{0}}A_{Y}(j_{1},j_{2})\nonumber\\
 & +\frac{\beta}{2}\epsilon^{p}\sum_{\left(k_{1},j_{1}\right)\in\gamma_{ZY}^{0}}\sum_{\left(k_{2},j_{2}\right)\in\gamma_{ZY}^{0}}A_{Y}(j_{1},j_{2})\nonumber\\
 & \leq\frac{\beta}{2}\epsilon^{p}\sum_{j_{1}\in U_{Y}\left(\gamma_{ZY}\right)}\sum_{j_{2}\in U_{Y}\left(\gamma_{ZY}\right)}A_{Y}(j_{1},j_{2})\nonumber\\
 & +\beta\epsilon^{p}\sum_{j_{1}\in U_{Y}\left(\gamma_{ZY}\right)}\sum_{\left(k_{2},j_{2}\right)\in\gamma_{ZY}^{0}}A_{Y}(j_{1},j_{2})\nonumber\\
 & +\frac{\beta}{2}\epsilon^{p}\sum_{\left(k_{1},j_{1}\right)\in\gamma_{ZY}^{0}}\sum_{\left(k_{2},j_{2}\right)\in\gamma_{ZY}^{0}}|A_{Y}(j_{1},j_{2})-A_{Z}(k_{1},k_{2})|\nonumber\\
 & +\frac{\beta}{2}\epsilon^{p}\sum_{k_{1}\in U_{Z}\left(\gamma_{XZ}\right)}\sum_{k_{2}\in U_{Z}\left(\gamma_{XZ}\right)}A_{Z}(k_{1},k_{2}). \label{eq:unassgined_edge_cost_Y}
\end{flalign}

\subsubsection{Combination of all edge costs}\label{subsec:combination of edge costs}
By combining the inequalities in \eqref{eq:assigned_edge_cost}, \eqref{eq:half-assigned_edge_cost_X}, \eqref{eq:half-assigned_edge_cost_Y}, \eqref{eq:unassigned_edge_cost_X}, and \eqref{eq:unassgined_edge_cost_Y}, we obtain
\begin{flalign}
     & e_{XY}^{p}\left(\gamma_{XY}\right)\nonumber \\
  \leq&\frac{\epsilon^{p}}{2}\sum_{\left(i_{1},k_{1}\right)\in\gamma_{XZ}^{1},}\sum_{\left(i_{2},k_{2}\right)\in\gamma_{XZ}^{1}}|A_{X}(i_{1},i_{2})-A_{Z}(k_{1},k_{2})|\nonumber \\
 & +\frac{\epsilon^{p}}{2}\sum_{\left(k_{1},j_{1}\right)\in\gamma_{ZY}^{1}}\sum_{(k_{2},j_{2})\in\gamma_{ZY}^{1}}|A_{Z}(k_{1},k_{2})-A_{Y}(j_{1},j_{2})|\nonumber \\
 & +\eta\epsilon^{p}\sum_{\left(i_{1},k_{1}\right)\in\gamma_{XZ}^{1}}\sum_{i_{2}\in U_{X}\left(\gamma_{XZ}\right)}A_{X}(i_{1},i_{2})\nonumber \\
 & +\eta\epsilon^{p}\sum_{\left(i_{1},k_{1}\right)\in\gamma_{XZ}^{1}}\sum_{\left(i_{2},k_{2}\right)\in\gamma_{XZ}^{0}}\text{|}A_{X}(i_{1},i_{2})-A_{Z}(k_{1},k_{2})|\nonumber \\
 & +\eta\epsilon^{p}\sum_{\left(k_{1},j_{1}\right)\in\gamma_{ZY}^{1}}\sum_{k_{2}\in U_{Z}\left(\gamma_{ZY}\right)}A_{Z}(k_{1},k_{2})\nonumber \\
& +\eta\epsilon^{p}\sum_{\left(k_{1},j_{1}\right)\in\gamma_{ZY}^{1}}\sum_{j_{2}\in U_{Y}\left(\gamma_{ZY}\right)}A_{Y}(j_{1},j_{2})\nonumber \\
 & +\eta\epsilon^{p}\sum_{\left(k_{1},j_{1}\right)\in\gamma_{ZY}^{1}}\sum_{\left(k_{2},j_{2}\right)\in\gamma_{ZY}^{0}}\text{|}A_{Y}(j_{1},j_{2})-A_{Z}(k_{1},k_{2})|\nonumber \\
 & +\eta\epsilon^{p}\sum_{\left(i_{1},k_{1}\right)\in\gamma_{XZ}^{1}}\sum_{k_{2}\in U_{Z}\left(\gamma_{XZ}\right)}A_{Z}(k_{1},k_{2})\nonumber \\
 & +\frac{\beta}{2}\epsilon^{p}\sum_{i_{1}\in U_{X}\left(\gamma_{XZ}\right)}\sum_{i_{2}\in U_{X}\left(\gamma_{XZ}\right)}A_{X}(i_{1},i_{2})\nonumber \\
 & +\beta\epsilon^{p}\sum_{i_{1}\in U_{X}\left(\gamma_{XZ}\right)}\sum_{\left(i_{2},k_{2}\right)\in\gamma_{XZ}^{0}}A_{X}(i_{1},i_{2})\nonumber \\
 & +\frac{\beta}{2}\epsilon^{p}\sum_{\left(i_{1},k_{1}\right)\in\gamma_{XZ}^{0}}\sum_{\left(i_{2},k_{2}\right)\in\gamma_{XZ}^{0}}|A_{X}(i_{1},i_{2})-A_{Z}(k_{1},k_{2})|\nonumber \\
 & +\frac{\beta}{2}\epsilon^{p}\sum_{k_{1}\in U_{Z}\left(\gamma_{ZY}\right)}\sum_{k_{2}\in U_{Z}\left(\gamma_{ZY}\right)}A_{Z}(k_{1},k_{2})\nonumber \\
 & +\frac{\beta}{2}\epsilon^{p}\sum_{j_{1}\in U_{Y}\left(\gamma_{ZY}\right)}\sum_{j_{2}\in U_{Y}\left(\gamma_{ZY}\right)}A_{Y}(j_{1},j_{2})\nonumber \\
 & +\beta\epsilon^{p}\sum_{j_{1}\in U_{Y}\left(\gamma_{ZY}\right)}\sum_{\left(k_{2},j_{2}\right)\in\gamma_{ZY}^{0}}A_{Y}(j_{1},j_{2})\nonumber \\
 & +\frac{\beta}{2}\epsilon^{p}\sum_{\left(k_{1},j_{1}\right)\in\gamma_{ZY}^{0}}\sum_{\left(k_{2},j_{2}\right)\in\gamma_{ZY}^{0}}|A_{Y}(j_{1},j_{2})-A_{Z}(k_{1},k_{2})|\nonumber \\
 & +\frac{\beta}{2}\epsilon^{p}\sum_{k_{1}\in U_{Z}\left(\gamma_{XZ}\right)}\sum_{k_{2}\in U_{Z}\left(\gamma_{XZ}\right)}A_{Z}(k_{1},k_{2}).\label{eq:substitution1}
\end{flalign}

Now, we operate again on the assigned edges and half-assigned edges.
For assigned edges, we have
\begin{flalign}
     & \frac{\epsilon^{p}}{2}\sum_{\left(i_{1},k_{1}\right)\in\gamma_{XZ},}\sum_{\left(i_{2},k_{2}\right)\in\gamma_{XZ}}|A_{X}(i_{1},i_{2})-A_{Z}(k_{1},k_{2})|\nonumber \\
  =&\frac{\epsilon^{p}}{2}\sum_{\left(i_{1},k_{1}\right)\in\gamma_{XZ}^{0}\cup\gamma_{XZ}^{1}}\sum_{\left(i_{2},k_{2}\right)\in\gamma_{XZ}^{0}\cup\gamma_{XZ}^{1}} \nonumber \\
 &|A_{X}(i_{1},i_{2})-A_{Z}(k_{1},k_{2})|\nonumber\\
 =&\frac{\epsilon^{p}}{2}\sum_{\left(i_{1},k_{1}\right)\in\gamma_{XZ}^{0}}\sum_{\left(i_{2},k_{2}\right)\in\gamma_{XZ}^{0}}|A_{X}(i_{1},i_{2})-A_{Z}(k_{1},k_{2})|\nonumber\\
& +\epsilon^{p}\sum_{\left(i_{1},k_{1}\right)\in\gamma_{XZ}^{0}}\sum_{\left(i_{2},k_{2}\right)\in\gamma_{XZ}^{1}}|A_{X}(i_{1},i_{2})-A_{Z}(k_{1},k_{2})|\nonumber\\
 & +\frac{\epsilon^{p}}{2}\sum_{\left(i_{1},k_{1}\right)\in\gamma_{XZ}^{1}}\sum_{\left(i_{2},k_{2}\right)\in\gamma_{XZ}^{1}}|A_{X}(i_{1},i_{2})-A_{Z}(k_{1},k_{2})|.
\end{flalign}
Following the conditions in Definition \ref{defn:definition of metric family}, it is met that $0 \leq \beta \leq 1$ and $0 \leq \eta \leq 1$. Therefore, the assigned parts in $X$ and $Z$ \eqref{eq:substitution1} have the following inequality:
\begin{flalign}
     & \frac{\epsilon^{p}}{2}\sum_{\left(i_{1},k_{1}\right)\in\gamma_{XZ}^{1},}\sum_{\left(i_{2},k_{2}\right)\in\gamma_{XZ}^{1}}|A_{X}(i_{1},i_{2})-A_{Z}(k_{1},k_{2})|\nonumber \\
 & +\eta\epsilon^{p}\sum_{\left(i_{1},k_{1}\right)\in\gamma_{XZ}^{1}}\sum_{\left(i_{2},k_{2}\right)\in\gamma_{XZ}^{0}}\text{|}A_{X}(i_{1},i_{2})-A_{Z}(k_{1},k_{2})|\nonumber \\
 & +\frac{\beta}{2}\epsilon^{p}\sum_{\left(i_{1},k_{1}\right)\in\gamma_{XZ}^{0}}\sum_{\left(i_{2},k_{2}\right)\in\gamma_{XZ}^{0}}|A_{X}(i_{1},i_{2})-A_{Z}(k_{1},k_{2})|\nonumber \\
  \leq &\frac{\epsilon^{p}}{2}\sum_{\left(i_{1},k_{1}\right)\in\gamma_{XZ}^{1},}\sum_{\left(i_{2},k_{2}\right)\in\gamma_{XZ}^{1}}|A_{X}(i_{1},i_{2})-A_{Z}(k_{1},k_{2})|\nonumber \\
 & +\epsilon^{p}\sum_{\left(i_{1},k_{1}\right)\in\gamma_{XZ}^{1}}\sum_{\left(i_{2},k_{2}\right)\in\gamma_{XZ}^{0}}\text{|}A_{X}(i_{1},i_{2})-A_{Z}(k_{1},k_{2})|\nonumber \\
 & +\frac{\epsilon^{p}}{2}\sum_{\left(i_{1},k_{1}\right)\in\gamma_{XZ}^{0}}\sum_{\left(i_{2},k_{2}\right)\in\gamma_{XZ}^{0}}|A_{X}(i_{1},i_{2})-A_{Z}(k_{1},k_{2})|\nonumber \\
  =&\frac{\epsilon^{p}}{2}\sum_{\left(i_{1},k_{1}\right)\in\gamma_{XZ},}\sum_{\left(i_{2},k_{2}\right)\in\gamma_{XZ}}|A_{X}(i_{1},i_{2})-A_{Z}(k_{1},k_{2})|.\label{eq:assigned_edge_inequality_XZ}
\end{flalign}
This result also applies to the assigned parts in $Z$ and $Y$ in \eqref{eq:substitution1}.
\begin{flalign}
    & \frac{\epsilon^{p}}{2}\sum_{\left(k_{1},j_{1}\right)\in\gamma_{ZY}^{1}}\sum_{(k_{2},j_{2})\in\gamma_{ZY}^{1}}|A_{Z}(k_{1},k_{2})-A_{Y}(j_{1},j_{2})|\nonumber \\
 & +\eta\epsilon^{p}\sum_{\left(k_{1},j_{1}\right)\in\gamma_{ZY}^{1}}\sum_{\left(k_{2},j_{2}\right)\in\gamma_{ZY}^{0}}\text{|}A_{Y}(j_{1},j_{2})-A_{Z}(k_{1},k_{2})|\nonumber \\
 & +\frac{\beta}{2}\epsilon^{p}\sum_{\left(k_{1},j_{1}\right)\in\gamma_{ZY}^{0}}\sum_{\left(k_{2},j_{2}\right)\in\gamma_{ZY}^{0}}|A_{Y}(j_{1},j_{2})-A_{Z}(k_{1},k_{2})|\nonumber \\
 & \leq\frac{\epsilon^{p}}{2}\sum_{\left(k_{1},j_{1}\right)\in\gamma_{ZY},}\sum_{\left(k_{2},j_{2}\right)\in\gamma_{ZY}}|A_{Y}(j_{1},j_{2})-A_{Z}(k_{1},k_{2})|.\label{eq:assigned_edge_inequality_ZY}
\end{flalign}

Then we consider the half-assigned edges between $X$ and $Z$ in \eqref{eq:substitution1}. Following the conditions in Definition \ref{defn:definition of metric family}, it is met that $\beta\leq \eta$. Then, we can write 
\begin{flalign}
     & \eta\epsilon^{p}\sum_{\left(i_{1},k_{1}\right)\in\gamma_{XZ}^{1}}\sum_{i_{2}\in U_{X}\left(\gamma_{XZ}\right)}A_{X}(i_{1},i_{2})\nonumber \\
 & +\eta\epsilon^{p}\sum_{\left(i_{1},k_{1}\right)\in\gamma_{XZ}^{1}}\sum_{k_{2}\in U_{Z}\left(\gamma_{XZ}\right)}A_{Z}(k_{1},k_{2})\nonumber \\
 & +\beta\epsilon^{p}\sum_{i_{1}\in U_{X}\left(\gamma_{XZ}\right)}\sum_{\left(i_{2},k_{2}\right)\in\gamma_{XZ}^{0}}A_{X}(i_{1},i_{2})\nonumber \\
  \leq&\eta\epsilon^{p}\sum_{\left(i_{1},k_{1}\right)\in\gamma_{XZ}^{1}}\sum_{i_{2}\in U_{X}\left(\gamma_{XZ}\right)}A_{X}(i_{1},i_{2})\nonumber \\
 & +\eta\epsilon^{p}\sum_{\left(i_{1},k_{1}\right)\in\gamma_{XZ}^{1}}\sum_{k_{2}\in U_{Z}\left(\gamma_{XZ}\right)}A_{Z}(k_{1},k_{2})\nonumber \\
 & +\eta\epsilon^{p}\sum_{i_{1}\in U_{X}\left(\gamma_{XZ}\right)}\sum_{\left(i_{2},k_{2}\right)\in\gamma_{XZ}^{0}}A_{X}(i_{1},i_{2})\nonumber \\
  =&\eta\epsilon^{p}\sum_{\left(i_{1},k_{1}\right)\in\gamma_{XZ}}\sum_{i_{2}\in U_{X}\left(\gamma_{XZ}\right)}A_{X}(i_{1},i_{2})\nonumber \\
 & +\eta\epsilon^{p}\sum_{\left(i_{1},k_{1}\right)\in\gamma_{XZ}^{1}}\sum_{k_{2}\in U_{Z}\left(\gamma_{XZ}\right)}A_{Z}(k_{1},k_{2})\nonumber \\
 \leq&\eta\epsilon^{p}\sum_{\left(i_{1},k_{1}\right)\in\gamma_{XZ}}\nonumber \\
 &\left[\sum_{i_{2}\in U_{X}\left(\gamma_{XZ}\right)}A_{X}(i_{1},i_{2})+\sum_{k_{2}\in U_{Z}\left(\gamma_{XZ}\right)}A_{Z}(k_{1},k_{2})\right]. \label{eq:half_XZ_inequality}
\end{flalign}
We now operate similarly on the half-assigned edges between $Z$ and $Y$, we have 
\begin{flalign}
     & \eta\epsilon^{p}\sum_{\left(k_{1},j_{1}\right)\in\gamma_{ZY}^{1}}\sum_{k_{2}\in U_{Z}\left(\gamma_{ZY}\right)}A_{Z}(k_{1},k_{2})\nonumber \\
 & +\eta\epsilon^{p}\sum_{\left(k_{1},j_{1}\right)\in\gamma_{ZY}^{1}}\sum_{j_{2}\in U_{Y}\left(\gamma_{ZY}\right)}A_{Y}(j_{1},j_{2})\nonumber \\
 & +\beta\epsilon^{p}\sum_{j_{1}\in U_{Y}\left(\gamma_{ZY}\right)}\sum_{\left(k_{2},j_{2}\right)\in\gamma_{ZY}^{0}}A_{Y}(j_{1},j_{2})\nonumber \\
  \leq&\eta\epsilon^{p}\sum_{\left(k_{1},j_{1}\right)\in\gamma_{ZY}} \nonumber \\
  &\left[\sum_{k_{2}\in U_{Z}\left(\gamma_{ZY}\right)}A_{Z}(k_{1},k_{2})+\sum_{j_{2}\in U_{Y}\left(\gamma_{ZY}\right)}A_{Y}(j_{1},j_{2})\right]\label{eq:half_ZY_inequality}.
\end{flalign}
If we substitute \eqref{eq:assigned_edge_inequality_XZ}, \eqref{eq:assigned_edge_inequality_ZY}, \eqref{eq:half_XZ_inequality} and \eqref{eq:half_ZY_inequality} into \eqref{eq:substitution1}, the inequality in \eqref{eq:edge_cost_inequality} is proved. 

\end{document}